\documentclass{ws-ijmpa}
\def\draftnote{~}
\def\be{\begin{equation}}
\def\ee{\end{equation}}
\def\refb#1{(\ref{#1})}
\def\laq{\raise 0.4ex\hbox{$<$}\kern -0.8em\lower 0.62ex\hbox{$\sim$}}
\def\gaq{\raise 0.4ex\hbox{$>$}\kern -0.7em\lower 0.62ex\hbox{$\sim$}}
\begin{document}
\markboth{Marco Cavagli\`a}{Black Hole and Brane Production in TeV Gravity}
\catchline{}{}{}
\title{BLACK HOLE AND BRANE PRODUCTION IN TEV GRAVITY: A REVIEW}
\author{\footnotesize MARCO CAVAGLI\`A\footnote{Email: marco.cavaglia@port.ac.uk}}
\address{Center for Theoretical Physics, Massachusetts Institute of Technology\\
77 Massachusetts Avenue, Cambridge MA 02139-4307, USA\\ {\rm and}\\ 
Institute of Cosmology and Gravitation, University of Portsmouth\\
Portsmouth PO1 2EG, U.K.\footnote{Present address.}}
\maketitle
\begin{abstract}
In models with large extra dimensions particle collisions with center-of-mass
energy larger than the fundamental gravitational scale can generate
non-perturbative gravitational objects such as black holes and branes. The
formation and the subsequent decay of these super-Planckian objects would be
detectable in particle colliders and high energy cosmic ray detectors, and have
interesting implications in cosmology and astrophysics. In this paper we
present a review of black hole and brane production in TeV-scale gravity. 
\end{abstract}

\section{Introduction}
In the spring of 1998 Arkani-Hamed, Dimopoulos and Dvali
(ADD)~\cite{Arkani-Hamed:1998rs,Arkani-Hamed:1998nn}, later joined by
Antoniadis~\cite{Antoniadis:1998ig}, proposed a possible solution to the
hierarchy problem of high-energy physics. In the ADD scenario the electroweak
scale ($E_{EW}\sim 1$ TeV) is identified with the ultraviolet cutoff of the
theory. Unification of electroweak and Planck scales takes place at $\sim$ 1
TeV. Gravity becomes strong at the electroweak scale and neither supersymmetry
nor technicolor are required to achieve radiative stability. The idea that the
gravitational scale can be lowered by some unknown physics dates back to the
early 90's when Antoniadis~\cite{Antoniadis:1990ew} first proposed that
perturbative string theories generally predict the existence of extra
dimensions at energies of order of the TeV scale. In presence of Large Extra
Dimensions~\footnote{Here ``large'' means larger than the fundamental scale.}
(LEDs) the observed weakness of gravity is a consequence of the ``leakage'' of
gravity in the extra dimensions. The Standard Model (SM) fields are constrained
in a four-dimensional submanifold of the higher-dimensional spacetime. The
success of the SM up to energies of a few hundreds of GeV indeed requires SM
fields to be localized on a three-brane embedded in the extra
dimensions~\cite{Rubakov:bb}. On the contrary, gravity and other non-SM fields
are allowed to freely propagate outside the three-brane.

\begin{figure}
\centerline{\psfig{file=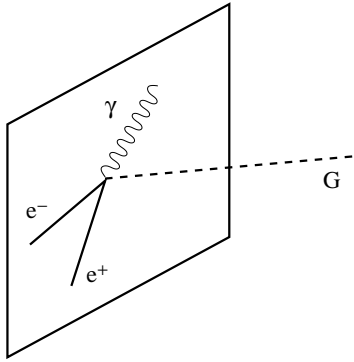,width=5cm}}
\vspace*{8pt}
\caption{A schematic illustration of the braneworld model. SM fields are
confined to the three-dimensional brane whereas gravitons can propagate in the
extra dimensions.}
\end{figure}

The recognition of the ADD scenario as a feasible model of high-energy physics 
was soon boosted by the possibility of including the ADD model in a consistent
theory of Quantum Gravity (QG): String Theory (ST). In the past years
ST~\cite{Polchinski:book} has emerged as the most successful candidate for a
consistent theory of QG. The most recent understanding of ST entails that the
five consistent STs and 11-dimensional supergravity (SUGRA) are connected by a
web of duality transformations. Type I, II and heterotic STs, and
11-dimensional SUGRA constitute special points of a large, multi-dimensional
moduli space of a more fundamental nonperturbative theory, called $M$-theory.
In addition to strings, the nonperturbative formulation of ST predicts the
existence of higher-dimensional, nonperturbative extended objects
($D$-branes)~\cite{Polchinski:book}. The presence of branes in the theory leads
naturally to the localization of the SM fields and to the hierarchy in the size
of the extra dimensions which are postulated in the ADD scenario~\footnote{In
the Ho\v rava-Witten model \cite{Horava:1996ma,Horava:1995qa}, for example, the
hierarchy of the extra dimensions and the confinement of the SM fields on a
three-brane follow from a double compactification of 11-dimensional SUGRA on a
one-dimensional orbifold and on a Calaby-Yau six-fold.}. Therefore, TeV-scale
gravity can be successfully realized in
ST~\cite{Antoniadis:2001sw,Benakli:1999yc,Shiu:1998pa}.

The attractiveness of the ADD proposal promptly opened up a whole new field of
investigation which is evident in the explosion of papers appeared in the
literature. Within months Dienes, Dudas and Gherghetta~\cite{Dienes:1998vh}
investigated the effects of extra dimensions on the GUT scale. Giudice,
Rattazzi and Wells~\cite{Giudice:1998ck}, and Mirabelli, Perelstein and
Peskin~\cite{Mirabelli:1998rt} studied the consequences of LEDs for high-energy
collider experiments. Han, Lykken and Zhang~\cite{Han:1998sg}, and
Hewett~\cite{Hewett:1998sn}, studied low-energy phenomenology of TeV-scale
Kaluza-Klein (KK) modes. Argyres, Dimopoulos and
March-Russell~\cite{Argyres:1998qn}, and Kaloper and Linde
\cite{Kaloper:1998sw}, investigated the properties of Black Holes (BHs) and
inflation in the LED scenario, respectively. Randall and Sundrum proposed an
alternative solution to the hierarchy problem based on the existence of warped
extra dimensions~\cite{Randall:1999ee,Randall:1999vf}. Afterward, the number of
papers on the subject has increased dramatically.

The presence of LEDs affects both sub- and super-Planckian physics.
Sub-Planckian physics is essentially affected by the presence of KK
modes~\cite{Han:1998sg,Hewett:1998sn}. The coupling of KK states to SM
particles leads to deviations from SM predictions in perturbative processes
such as quarkonium radiative decays, two- and four-fermion interactions and
production of gauge bosons. In the LED scenario effective interactions
resulting from massive KK modes are suppressed by powers of the fundamental
Planck mass $M_\star\sim O({\rm TeV})$. This leads to significant experimental
signatures. Super-Planckian physics~\footnote{The physics of super-Planckian
collisions has a long history dating back to the late 80's --- early 90's. See
Refs.~\cite{'tHooft:rb,Amati:1987wq,Amati:1987uf,Verlinde:1991iu}.} involves
non-perturbative effects. The most striking super-Planckian phenomenon is the
formation of Non-Perturbative Gravitational Objects (NPGOs) from
hard-scattering processes. Creation of BHs from super-Planckian scattering was
advocated by Banks and Fischler~\cite{Banks:1999gd} soon after the appearance
of the papers on non-perturbative sub-Planckian effects. Particle collisions
with Center-of-Mass (CM) energy larger than the fundamental Planck mass, and
impact parameter smaller than the Schwarzchild radius associated with the CM
energy, are expected to generate BHs. A completely inelastic process creates a
single BH which subsequently decays via Hawking radiation~\cite{Hawking:sw}
and can be regarded as an intermediate metastable state in the super-Planckian
scattering. BH formation is expected to dominate hard-scattering processes
because the number of non-perturbative states grows faster than the number of
perturbative string states. Moreover, the cross-section for BH formation grows
with energy faster than cross-sections of non-perturbative sub-Planckian
processes. Formation and evaporation of super-Planckian BHs would be detectable
in future hadron
colliders~\cite{Giddings:2001bu,Dimopoulos:2001hw,Cheung:2001ue,Uehara:2002cj,Uehara:2002gv,Anchordoqui:2002cp}
such the Large Hadron Collider (LHC)~\cite{LHC} or the Very Large Hadron
Collider (VHLC)~\cite{VLHC}. Ultra High-Energy Cosmic Rays (UHECRs) impacting
Earth's atmosphere with CM energy of a few hundreds of TeV~\cite{Nagano:ve}
could also produce transient BHs~\cite{Feng:2001ib} that would account for
features in the spectrum of UHECR~\cite{Kazanas:2001ep,Anchordoqui:2001ei} and
be observed in neutrino
telescopes~\cite{Uehara:2001yk,Kowalski:2002gb,Alvarez-Muniz:2002ga}, ground
array and air fluorescence
detectors~\cite{Ringwald:2001vk,Anchordoqui:2001cg,Anchordoqui:2002fc}.~\footnote{See
Refs.~\cite{Anchordoqui:2002hs,Sigl:2002yk} for recent and comprehensive
reviews.}

If super-Planckian collisions create NPGOs, production of spherically symmetric
BHs is just the simplest of a plethora of possible super-Planckian physical
processes. At Planckian or super-Planckian energies, we also expect creation of
other NPGO which are predicted by QG theories. 

$p$-branes are $p$-dimensional spatially extended solutions of gravitational
theories in higher dimensions, such as higher-dimensional Einstein-Maxwell
theory~\cite{Gregory:1995qh,Cavaglia:1997hc} or low-energy effective
STs~\cite{Duff:1993ye,Duff:1996hp,Stelle:1996tz}. They are translationally
invariant in $p$-dimensions and isotropic in the $D-p-1$ spatial directions
transverse to the translationally-symmetric ones, i.e., $p$-branes preserve the
$\hbox{(Poincar\'e)}_{p+1}\times {\rm SO}(D-p-1)$ symmetry. Therefore, they can
be viewed as Poincar\'e-invariant hyperplanes propagating in spacetime.
$p$-branes are characterized by the tension $T$, the mass $M_p$ and a set of
conserved charges. They possess naked singularities and/or horizons. In the
former case $p$-brane solutions are interpreted as the exterior metric of
higher-dimensional extended objects such as (cosmic)
strings~\cite{Gregory:1995qh}. In the latter, they are usually called ``black''
branes because their geometry is that of a $(N-p)$-dimensional hole $\times$
$p$-dimensional hyperplane. A $0$-brane is a spherically symmetric BH. Branes
are a generic phenomenon of any gravitational theory. The presence of extra
dimensions is indeed sufficient to permit the existence of these extended
objects, though their physical properties depend on the theory. Therefore, if
the fundamental Planck scale is of order of TeV, it is reasonable to expect
creation of branes at energies above the TeV
scale~\cite{Ahn:2002mj,Ahn:2002zn}. Super-Planckian physics is affected by the
production of BHs and $p$-branes alike. TeV-branes can produce experimental
signatures in particle colliders~\cite{Ahn:2002mj,Ahn:2002zn,Cheung:2002aq} and
in UHECR
physics~\cite{Ahn:2002mj,Ahn:2002zn,Jain:2002kf,Anchordoqui:2002it,Anchordoqui:2002hs}.
In cosmology, primordial creation of stable branes may have played an important
role in the dynamics of the very early universe when the temperature was above
the TeV scale; brane relics could be the observed dark
matter~\cite{Ahn:2002zn}. 

The purpose of this article is to review the theory and the phenomenology of BH
and brane production in LED gravity and present an up-to-date bibliography on
the subject. In the next section we will review the basic features of TeV-scale
gravity and LED models. The physics of brane formation in TeV-scale gravity
follows closely that of BHs. Therefore, we first discuss BH creation in LED
models in Sect.~3. Sections 4-7 deal with TeV-branes. In Sect.~4 we review
brane solutions in Einstein-Maxwell and low-energy STs and discuss their
formation and basic properties. Sections 5-7 deal with experimental signatures
in particle colliders, UHECRs and cosmology, respectively. Conclusion and
outlook complete the review.
\section{Hierarchy Problem and TeV Gravity}
Two different fundamental energy scales are observed in nature: The electroweak
scale $E_{EW}\sim 1$ TeV and the gravitational scale $E_{G}\sim 10^{16}$ TeV.
The SM of particles successfully explains particle physics up to the $E_{EW}$
energy. However, little is known about quantum physics above the electroweak
scale. One of the most challenging issues is to explain the hierarchy problem,
{\it i.e.} the largeness and radiative stability of the ratio $E_{G}/E_{EW}$.
\subsection{Gravity and Extra Dimensions}
Classical gravity is described by the Einstein-Hilbert action
\be
S_{EH}={1\over 16\pi G_4}\int_{\cal M} d^4x \sqrt{-g} {\cal R}(g)\,,
\label{action-Einstein}
\ee
where $\cal M$ is a four-dimensional hyperbolic manifold with metric
$g_{\mu\nu}$, $g$ is the determinant of the metric, ${\cal R}(g)$ is the Ricci
scalar, and $G_4 = 6.673(10)\cdot 10^{-11}~{\rm m}^3~{\rm kg}^{-1}~{\rm s}^{-2}$
= $6.707(10)\cdot 10^{-39}~\hbar c~({\rm GeV}/c^2)^{-2}$ is the gravitational
constant (Newton's constant). In natural units $G_4$ has dimensions of
inverse mass squared. The Planck mass
\be
M_{\rm Pl}\equiv G_4^{-1/2}\sim 1.22\cdot 10^{16}~{\rm TeV}
\label{MPl}
\ee
determines the gravitational scale at which quantum gravitational phenomena
become strong. At low energy, gravity manifests itself as
a long-range attractive force with coupling constant $G_4$. The gravitational
potential of a massive object with mass $M$ at a distance $r$ is
\be
V(r)=-G_4{M\over r}\,.
\label{g4pot}
\ee
Experiments with a torsion pendulum/rotating attractor
instrument~\cite{Hoyle:2000cv,Adelberger:2002ic} have recently tested the
gravitational force at length scales below 1 mm without evidence for violations
of Eq.~\refb{g4pot}. 

Any consistent theory of QG must reduce to the Einstein-Hilbert theory in the
low-energy limit. ST requires that we live in a higher-dimensional spacetime.
In the Einstein frame, the generic bosonic sector of low-energy effective STs
is described by the action
\be
S={1\over 16\pi G_D}\int D^Dx\sqrt{-g}\left[{\cal R}(g)-a(\nabla\phi)^2
-{1\over 2n!}e^{b\phi}F^2_{[n]}+{\rm CS~terms}\right]\,.
\label{action-strings}
\ee
where $\phi$ is a scalar field (dilaton), $F_{[n]}$ is the field strength of
the $(n-1)$-form gauge potential $A_{[n-1]}$ and $D=10$ or 11. The parameters
$a$, $b$ and $n$, and the Chern-Simons (CS) terms, depend on the model. For
instance, 11-dimensional SUGRA has $a=b=0$, $n=4$ and CS terms equal to
${1\over 6}F_{[4]}\wedge F_{[4]}\wedge A_{[3]}$. If ST is the
ultimate theory of QG, six or seven dimensions must be compactified
to yield a four-dimensional effective action. This is generally accomplished
by assuming that the higher-dimensional spacetime is a (warped) product of
four-dimensional Minkowski spacetime and a compact $(D-4)$-dimensional
Riemannian manifold. The spacetime metric is
\be
ds^2 = g_{ab}(x)dx^a dx^b=
e^{2A(y)}dx^\mu dx^\nu\eta_{\mu\nu}+h_{ij}(y)dy^idy^j\,,
\label{gen-metric}
\ee
where the Greek indices run from $0$ to $3$ and the coordinates $y$
($i,j=4\dots D-1$) parametrize the compactified dimensions. In such a scenario,
the observed Planck scale $M_{\rm Pl}$ is a quantity derived from the
$D$-dimensional fundamental Planck scale~\footnote{For notations see Appendix
A.}
\be
M_\star\equiv G_D^{-1/(D-2)}\,.
\label{mstar}
\ee
The four-dimensional Newton's constant is related to the $D$-dimensional
gravitational constant $G_D$ by the relation
\be
G_4={G_D\over V_{D-4}}\,,\qquad\to\qquad M_{\rm Pl}^2=
{M_\star^{D-2} V_{D-4}}\,,
\label{G4}
\ee
where
\be
V_{D-4}=\int d^{D-4}y\sqrt{h}\,e^{2A(y)}
\label{volume}
\ee
is the volume of the extra-dimensional space modulated by the warp factor
$e^{A(y)}$. The ratio of the fundamental Planck constant $M_\star$ to the
observed Planck constant $M_{\rm Pl}$ depends on the geometry and on the size
of the compactified space. If the volume of the latter is of order of the
fundamental Planck scale, then all quantities in Eq.~\refb{G4} are of order
$\sim 10^{16}$ TeV. However, other scenarios are possible: In the Ho\v
rava-Witten model, for instance, $M_\star$ is of order of the GUT scale.

Equation \refb{G4} provides a possible solution to the hierarchy problem, which
is bypassed by identifying $M_\star$ with the electroweak scale. The volume of
the extra-dimensional space is large in fundamental Planck units. For instance,
assuming for simplicity a symmetric toroidal compactification with radii
$R_i=L/2\pi$ the condition $M_\star\sim M_{\rm EW}$ gives the relation:
\be
L \sim 10^{30/n - 17} {\rm cm} \times \left(
{\rm TeV}\over {M_{\rm EW}}\right)^{1+2/n}\,,
\ee
where $n=D-4$ is the number of extra dimensions. An upper limit to the size of
the compactified space can be obtained by measuring the gravitational potential
at small distances. Indeed, if $n$ extra dimensions open up at the scale $L$,
the gravitational potential at scales smaller than $L$ behaves as
\be
V(r)\sim -G_{n+4}{M\over r^{n+1}}\,.
\label{gnpot}
\ee
Clearly, the non-observation of deviations from the four-dimensional behavior
of Eq.~\refb{g4pot} at a distance $\sim L'$ constrains the size of the extra
dimensions and the fundamental Planck scale to be larger than $L'$ and
${L'}^{-1}$, respectively. The latest experimental
results~\cite{Adelberger:2002ic} suggest that the gravitational force follows
the inverse square law up to distances of 150 $\mu$m. For a two-dimensional
symmetric toroidal compactification this implies a fundamental Planck scale
$M_\star\,\gaq\, 1.6$ TeV.
\subsection{Theoretical Models}
A lot of efforts have been devoted to the formulation of LED models in ST and
M-theory (see, e.g., Ref.~\cite{Antoniadis:1999fj} for a short review).
Although the LED scenario does not require ST, the two main ingredients of TeV
gravity, namely the hierarchy of scales and the confinement of the SM fields on
a three-branes, can be easily accommodated in ST. 
\subsubsection{Heterotic String Theory}
In weakly coupled heterotic ST, the relation between the four-dimensional
Planck scale and the string scale $M_s$ is 
\be
M_{\rm Pl}^2={1\over \lambda_s^2}M_s^8V_6={1\over g^2}M_s^2\,,
\label{Ms}
\ee
where $V_6$ is the volume of the compactified space,
$\lambda_s=e^{\langle\phi\rangle}\ll 1$ is the vacuum expectation value of the
dilaton field, and $g$ is the gauge coupling. Since $g\sim 1/5$ the heterotic
string scale appears to be near the Planck scale. The string is weakly coupled
when the compactified volume is of order of the string scale. However, in order
to break supersymmetry by compactification at a scale smaller than the
heterotic string scale, a large compactification radius is needed. In this case
$\lambda_s$ becomes large and ST is strongly coupled. Strongly coupled ST can
be discussed by means of dualities. For instance, strongly coupled ST with one
(two or more) large radii is given by weakly coupled type IIB (IIA or I/I') ST.
In the strongly coupled regime Eq.~\refb{Ms} is no more valid. The string
tension becomes an arbitrary parameter which can provide a solution to the
hierarchy problem.
\subsubsection{Type I/I' String Theory}
A consistent framework for realizing the ADD scenario is type I ST. The
strongly coupled regime of $SO(32)$ heterotic ST is given by type I or I' ST.
In type I/I' ST closed and open strings describe gravity and gauge fields,
respectively. The SM fields are confined on a collection of $Dp$-branes, where
the ends of the open strings are constrained to
propagate~\cite{Antoniadis:1998ig,Shiu:1998pa}. The Planck scale and the gauge
coupling are
\be
M_{\rm Pl}^2={1\over\lambda_s^2}M_s^8V_{L}V_{T}\,,\qquad
g^{-2}={1\over\lambda_s}M_s^{p-3}V_{L}\,,
\label{MsI}
\ee
where $V_{L}$ and $V_{T}$ are the volume of the $(p-3)$-dimensional
longitudinal space and the volume of the $(9-p)$-dimensional transverse space to the
$Dp$-brane, respectively. Calculability at perturbative level ($\lambda_s<1$)
requires that  the $p-3$ longitudinal dimensions are compactified on the string
scale. For a symmetric transverse compactification with radius $R_{T}$,
Eq.~\refb{MsI} gives
\be
M_{\rm Pl}^2={1\over g^4v_{L}}M_s^{11-p}R_{T}^{9-p}\,,
\label{MsIb}
\ee
where $v_{L}\sim 1$ is the longitudinal volume in string units. If the $9-p$
transverse dimensions are compactified on a much larger scale than the string
scale $M_s$, the string scale can be smaller than the Planck scale. Therefore,
gravity becomes strong at a much lower scale ($M_s$) than the Planck scale,
though the string is weakly coupled. 
\subsubsection{Type IIA/IIB String Theory}
Heterotic ST compactified to six or less dimensions admits a dual description
as a type II ST. The gauge coupling is independent from the string coupling
$\lambda_s$. Let us consider for simplicity a compactification on $K3\times
T^2$. For type IIA ST we have
\be
M_{\rm Pl}^2={1\over\lambda_s^2}M_s^2v_{K3}v_{T^2}\,,\qquad
g^{-2}=v_{T^2}\,,
\label{MsIIA}
\ee
where $v_{K3}$ and $v_{T^2}$ are the volume of the Calaby-Yau and of the
two-torus in string units, respectively. The second relation implies
$v_{T^2}\sim 1$. The Planck scale is
\be
M_{\rm Pl}^2={1\over g^2}{v_{K3}\over \lambda_s^2}M_s^2\,.
\label{MsIIAb}
\ee
In the previous equation the volume of the Calabi-Yau manifold and the string
coupling are free parameters. We can lower the string scale either by choosing
a large $K3$ volume or a small string coupling $\lambda_s\sim
10^{-14}$~\cite{Antoniadis:1999rm,Lykken:1996fj}. In the latter scenario,
gravity remains weak up to the Planck scale. Therefore, there are no obseravble
quantum gravitational effects at the TeV scale.

For type IIB ST we have
\be
M_{\rm Pl}^2={1\over\lambda_s^2}M_s^2v_{K3}v_{T^2}\,,\qquad
g^{-2}={R_1\over R_2}\,,
\label{MsIIB}
\ee
where $R_1$ and $R_2$ denote the radii of the two-dimensional torus. Setting
$R_2=g^2 R_1$, the relation between the Planck scale and the string scale is
\be
M_{\rm Pl}={g\over\lambda_s}v_{K3}^{1/2}R_1 M_s^2\,.
\label{MsIIBb}
\ee
The largest value for the string scale $M_s\sim 10^8$ TeV is obtained by
choosing $R\sim$ TeV and $v_{K3}\sim 1\sim\lambda_s$.
\subsubsection{M-Theory}
Strongly coupled $E_8\times E_8$ ST compactified on a Calabi-Yau six-fold
$CY_6$ is described by 11-dimensional SUGRA compactified on $S^1/Z_2\times
CY_6$. It follows that 
\be
M_{\rm Pl}^2=V_{CY} R_{11}M_s^9\,,\qquad g^{-2}=V_{CY}M_s^6\,,
\ee
where $V_{CY}$ is the volume of the Calabi-Yau six-fold and $R_{11}$ is the
radius of the orbifold. Setting $M_s\sim 1$ TeV we obtain $R_{11}\sim 10^{11}$
m, which is excluded experimentally. The lowest value of the M-theory scale
which is experimentally admissible is $M_s\sim 10^4$ TeV, corresponding to
$R_{11}\sim 1$ mm.

\begin{table}[h]
\tbl{Possible LED scenarios in ST/M-theory. The first two columns
give the number of LEDs and Small Extra Dimensions (SEDs) $\sim M_s^{-1}$. The
third column gives the string/M-theory scale $M_s$.}
{\begin{tabular}{@{}c@{\hbox to 12mm{\hfill}}c@{\hbox to 12mm{\hfill}}c@{\hbox
to 12mm{\hfill}}c@{}} \toprule
&LED & SED & $M_s$ \\ \colrule
I/I' ST & $n\ge 2$ (fm -- mm) & $6-n$ & $\sim$ TeV \\
IIA ST & $2\le n\le 4$ (fm -- mm)& $6-n$ & $\sim$ TeV \\
IIB ST & 2 ($\sim $ fm) & 4 & $\sim
10^{8}$ TeV \\
M-theory & 1 ($\laq$ mm) & 6 & $\gaq 10^4$ TeV \\ \botrule
\end{tabular}}
\end{table}

\subsubsection{String Theory and Warped Metrics}
Warped metrics arise naturally in STs thanks to the presence of
$D$-branes. (See, e.g., Refs.~\cite{Antoniadis:2001bh,Forste:2001ah} for recent
reviews.) A simple example of $D$-brane--induced warping has been given by
Verlinde~\cite{Verlinde:1999fy} who considers a stack of $N$ $D3$-branes with
geometry $AdS_5\times S^5$, where the $AdS_5$ submanifold is represented as
the warped product of a Poincar\'e-invariant four-dimensional spacetime and a
radial direction. The Verlinde model has recently been extended in
Ref.~\cite{Giddings:2001yu}. 
\subsection{Phenomenology of Compactification Models}
In this section we briefly discuss a few simple examples of compactification.
At a phenomenological level, the $n$-dimensional internal space is
characterized by the existence of one or more compactification scales. The
metric of the spacetime is given in Eq.~\refb{gen-metric}. Compactifications
are divided in two categories: Factorizable ($A(y)=0$) and non-factorizable or
``warped'' ($A(y)\not=0$).  
\subsubsection{Toroidal Compactifications}
The simplest example of factorizable compactification is the $n$-torus
($h_{ij}=R_i\delta_{ij}$, $y_i\in[0,2\pi[$). If all the radii $R_i$ have the
same size $R$, the toroidal compactification is called symmetrical. The ratio
between the Planck scale and the fundamental scale is:
\be
\left({M_{\rm Pl}\over M_{\star}}\right)^{2}=
\left({L'\over L_{\star}}\right)^{n}\equiv {l'}^n\,,
\label{symm-compact}
\ee
where $L'=2\pi R$ and $L_\star=M_\star^{-1}$. ST seems to favor the existence of
more than one compactification scale. In the simplest scenario with $m<n$ extra
dimensions compactified on the $L$ scale and $n-m$ dimensions compactified on
the $L'$ scale we have
\be
\left({M_{\rm Pl}\over M_{\star}}\right)^{2}=l^{m}l'^{n-m}\,.
\label{asymm-compact}
\ee
If $L\sim L_\star$, Eq.~\refb{asymm-compact} reduces to
Eq.~\refb{symm-compact} with $n-m\to n$. Therefore, experimental constraints
for the $n$-dimensional symmetric model apply also to the asymmetric model with
$n-m$ large dimensions. Asymmetric compactifications in the LED context were
first discussed by Lykken and Nandi~\cite{Lykken:1999ms}.
\subsubsection{Fat Branes and Universal Extra Dimensions}
In the compactification models considered above, the SM particles are
constrained to propagate on three-dimensional branes of infinitesimal
thickness. However, for asymmetric compactifications with small extra
dimensions, we can think of a scenario in which the SM particles propagate in
spatial dimensions other than the macroscopic ones. 

In the Fat Brane (FB) model the three-brane can be thought as a thick wall,
with the SM particles propagating inside the wall. The FB model was first
discussed by Arkani-Hamed and Schmaltz (AS) in Ref.~\cite{Arkani-Hamed:1999dc}.
In the AS scenario the wall is thick in one dimension. The Higgs and the SM
gauge fields are free to propagate inside the wall whereas the SM fermions are
stuck at different depths in the wall. Therefore, fermions and gauge fields
``see'' three and four spatial dimensions, respectively. If the wall thickness
is of order of the TeV$^{-1}$ scale, the AS model leads to interesting
perturbative effects that would be detectable at future particle colliders.

In the Universal Extra Dimension (UED) scenario~\cite{Appelquist:2000nn} all SM
particles are allowed to propagate freely in some of the extra dimensions,
provided that these dimensions are smaller than a few hundreds of GeV$^{-1}$. A
peculiar feature of the UED model is that the KK excitations are produced in
groups of two or more. This leads to different signatures in particle collider
experiments.
\subsubsection{Randall-Sundrum Compactification}
The simplest example of a warped compactification is the Randall-Sundrum (RS)
model~\cite{Randall:1999ee,Randall:1999vf}. The RS spacetime is
five-dimensional. The metric is
\be
ds^2 = e^{-2kr_c|y|}\eta_{\mu\nu}dx^\mu dx^\nu+r_c^2dy^2\,,
\label{RS-metric}
\ee
where $k$ is a parameter of order of the fundamental Planck scale and $r_c$ is
the radius of the finite-size extra dimension (orbifold) $y$ ($0\le y\le\pi$).
The RS metric \refb{RS-metric} is a solution of five-dimensional Einstein
gravity coupled to a negative cosmological constant and to two three-branes
located at the fixed points of the orbifold:
\be
S={M_\star^3\over 16\pi}\int
dx^4dy\sqrt{-g}\left(R-\lambda\right)+\sum_{i=1,2}\int dx^4
\sqrt{-h^{(i)}}\left({\cal L}^{(i)}-V^{(i)}\right)\,,
\ee
where $h_{\mu\nu}^{(i)}$, ${\cal L}^{(i)}$ and $V^{(i)}$ are the
four-dimensional metric, Lagrangian density, and vacuum energy of the $i$-th
brane, respectively. The bulk spacetime is a slice of the five-dimensional
Anti-de Sitter (AdS) geometry with curvature $\lambda$. The bulk cosmological
constant and the brane vacuum energy depend on the scale $k$:
\be
V_1=-V_2={3\over 4\pi}M_\star^3 k\,,\qquad\lambda=-12k^2\,.
\ee
Thus the branes have opposite tension. The scale $k$ is smaller than the
fundamental scale, i.e., the AdS curvature is small in fundamental units. The
radius of the orbifold is small but larger than $k^{-1}$ so that $r_c k\gg 1$.
Using Eqs.~\refb{G4} and \refb{volume} we find that the relation between the
observed Planck mass and the fundamental Planck mass is
\be
M_{\rm Pl}^2={M^3_\star\over k}\left[1-e^{-2kr_c\pi}\right]\sim {M^3_\star\over k}\,.
\ee
The scale of the physical phenomena on the branes is fixed by the value of the warp
factor. On the brane located at $y=\pi$ (visible brane), the conformal factor
of the metric is $\rho^2=e^{-2kr_c\pi}$ and all physical masses are rescaled
by a factor $\rho$. If $kr_c\sim 12$ physical mass scales of order of TeV are
generated from a fundamental Planck scale $\sim 10^{16}$ TeV. Alternatively,
the TeV scale can be regarded as the fundamental scale and the Planck mass as
the derived scale.

The second RS model~\cite{Randall:1999vf} is obtained from Eq.~\refb{RS-metric}
by letting $y_c\to\infty$. The negative tension brane is located on the AdS
horizon and the model formally contains a single brane.
\subsection{Experimental Constraints}
The size of the extra dimensions and the value of the fundamental Planck scale
are constrained by experiments. The constraints for non-warped symmetric
toroidal compactifications have been extensively studied in the
literature.~\footnote{For recent reviews see, e.g.,
Refs.~\cite{Peskin:2000ti,Uehara:2002yv,Cheung:1999fj}.} These constraints can
immediately be extended to the asymmetric toroidal compactification with
smaller scale of order of the fundamental Planck length.

We have already seen that experiments with a torsion
pendulum~\cite{Hoyle:2000cv,Adelberger:2002ic} limit the size of $n=2$ LEDs to
150 $\mu$m and the value of $M_\star$ to be larger than $1.6$ TeV. A lower
bound on the value of the fundamental Planck scale can also be derived from
particle collider experiments and astrophysical and cosmological
considerations. 
\subsubsection{Particle Collider Experiments}
The particle collider experiments are divided in two categories: non-perturbative
and perturbative. The former involves creation of super-Planckian NPGOs such as
BHs~\cite{Banks:1999gd} and branes~\cite{Ahn:2002mj}, and will be discussed in
the following sections. The latter consists essentially in missing-energy
experiments (due to emission of a real graviton) or search for deviations from
SM predictions in fermion-fermion interactions (due to virtual graviton
exchange)~\cite{Giudice:1998ck,Peskin:2000ti,Cheung:1999fj}. In missing-energy
experiments, the idea is to look for processes in which a graviton is scattered
off the brane, as in Fig.~1. The deviations from the SM expectations due to the missing
momentum carried with the graviton are model-independent~\cite{Cullen:2000ef}
and can be measured. The simplest processes of this kind
are~\cite{Peskin:2000ti}
\be
e^+ e^-\to\gamma G\,,\qquad q\bar q\to gG\,,
\ee
where $G$ is a graviton. In the first process, the smoking gun would be the
observation of a single photon with missing energy in a
$e^+e^-\to\nu\bar\nu\gamma(\gamma)$
background~\cite{Acciarri:1999kp,Cheung:1999fj}. The current and future lower bounds
on the fundamental Planck scale due to real graviton emission have been
given by Cullen {\it et al} in Ref.~\cite{Cullen:2000ef} (see Table 2).

\begin{table}[h]
\tbl{Lower bounds (95\% c.l.) on the fundamental Planck scale $M_\star$ (TeV)
from missing-energy experiments at present (LEP, Tevatron) and future (LC, LHC)
particle colliders for $2$- $4$- and $6$-dimensional symmetric toroidal
compactifications.}
{\begin{tabular}{@{}l@{\hbox to 25mm{\hfill}}c@{\hbox to 25mm{\hfill}}c@{\hbox
to 25mm{\hfill}}c@{}} \toprule 
&n=2 & n=4 & n=6 \\ \colrule
Lep II& 0.90 & 0.33 & 0.18 \\
Tevatron & 0.86 & 0.39 & 0.27 \\
LC & 5.8 & 2.0 & 1.1 \\
LHC & 9.4 & 3.4 & 2.1 \\
\botrule
\end{tabular}}
\end{table}

The limits on the fundamental Planck scale from virtual graviton effects have
been given by L3~\cite{Acciarri:1999jy}, ALEPH~\cite{ALEPH},
DELPHI~\cite{Abreu:2000ap}, OPAL~\cite{Abbiendi:1999wm},
H1~\cite{Adloff:2000dp} and D0~\cite{Abbott:2000zb} collaborations for various
processes. The most stringent bound comes from the D0 experiment and is given
in Table 3.
\begin{table}[h]
\tbl{Lower bounds (95\% c.l.) on the fundamental Planck scale $M_\star$ (TeV)
from virtual graviton exchange processes for $n$-dimensional symmetric
toroidal compactifications ($D0$ collaboration).}
{\begin{tabular}{@{}c@{\hbox to 11mm{\hfill}}c@{\hbox to 11mm{\hfill}}
c@{\hbox to 11mm{\hfill}}c@{\hbox to 11mm{\hfill}}c@{\hbox to 11mm{\hfill}}
c@{\hbox to 11mm{\hfill}}c@{}} \toprule
n & 2 & 3 & 4 & 5 & 6 & 7\\ \colrule
$M_\star$ & 0.88 & 0.77 & 0.58 & 0.47 & 0.39 & 0.35\\
\botrule
\end{tabular}}
\end{table}

\subsubsection{Astrophysical and Cosmological Constraints}
The astrophysical constraints on $M_\star$ are derived from the study of
supernova cooling and neutron star heat excess. In the cooling process of
type-II supernovae, the SM predicts that most of the energy is radiated through
neutrino emission. In the LED scenario, the KK modes created through
gravi-bremsstrahlung processes can also carry energy away. In order to avoid
overcooling, the fundamental Planck scale must be not too small. Lower bounds
on $M_\star$ have been estimated by Cullen and Perelstein in
Ref.~\cite{Cullen:1999hc} (and subsequently refined by several other authors in
Refs.~\cite{Barger:1999jf,Hanhart:2001fx,Hannestad:2001jv}). Using data from
the supernova SN1987A, Cullen and Perelstein give a lower bound of
$M_\star\gaq$ 38, 2.2, and 0.45 TeV for $n=2$, 3, and 4 extra dimensions,
respectively.~\footnote{Hannestad and Raffelt \cite{Hannestad:2001jv} give the
more restrictive bounds $M_\star\gaq$ 63 (3.9) TeV for $n=2$ (3).} The limits
from supernova cooling are much stronger than the limits from collider
experiments for $n=2,3$ but become unimportant for higher
$n$~\cite{Cullen:2000ef,Peskin:2000ti}. However, it should be stressed that the
results based on supernova cooling are affected by large uncertainties on the
temperature and on the density of the stellar core at collapse. The bounds on
$M_\star$ from neutron star heat excess are estimated by constraining the
heating process of neutron stars due to the decay and the subsequent absorption of
KK gravitons~\cite{Hannestad:2001xi}. The lower limit on $M_\star$ is $\sim
1260$ ($33$) TeV for $n=2$ (3). This result is also affected by large
theoretical and experimental uncertainties. 

The cosmological constraints on $M_\star$ are derived from the physics of the
Cosmic Microwave Background Radiation (CMBR) and of the Cosmic Gamma-Ray
Background Radiation (CGRBR). In the former, a lower bound on $M_\star$ is
estimated by constraining the cooling rate of the CMBR. The presence of
primordial KK gravitons increases the amount of matter in the universe, leading
to a more rapid cooling~\cite{Fairbairn:2001ct}. This sets the lower bounds
$M_\star\gaq$ 65 -- 750,  4 -- 32 and 0.7 -- 4 for $n=2$, 3 and 4,
respectively. The presence of KK gravitons leads to modifications in the CGRBR.
Primordial KK modes can decay in photons and affect the CGRBR. This sets a
bound on the primordial nucleosynthesis normalcy temperature which leads to
$M_\star\gaq\, 83$ -- 263 (2.8 -- 7.6) TeV for two (three)
LEDs~\cite{Hall:1999mk}. Constraints derived from overclosure of the universe
due to KK modes are generally weaker.

Summarizing, in the toroidal compactification scenario particle collider
experiments and astrophysical and cosmological observations can be used to set
a lower bound on the fundamental Planck scale. The constraints from
astrophysics and cosmology dominate over the constraints derived from particle
collider experiments for $n=2,3$ LEDs, though they are affected by larger
uncertainties. The estimated bounds on $M_\star$ seem to rule out a scenario
with two (and possibly three) large extra dimensions. Weaker constraints from
particle collider experiments dominate for higher-dimensional
compactifications.
\section{Black Holes and TeV Gravity}
In the previous section we have discussed the experimental signatures of LEDs in
sub-Planckian physics. In this regime gravity is perturbative. Low-scale
gravity manifests itself as KK modes which couple to SM fields. When energies
reach the fundamental Planck scale, gravity becomes non-perturbative and
quantum gravitational effects become strong. ST is believed to describe
super-Planckian gravitational processes. Thus it is reasonable to expect that
super-Planckian particle collisions produce NPGOs which are predicted by ST,
such as BHs, string balls~\cite{Dimopoulos:2001qe}, and branes. In the LED
scenario this occurs for processes with energy of order of the TeV scale. At
energies well above the Planck scale gravity becomes semiclassical. The
entropy of the processes is $S\gg 1$ and gravitational interactions are described
by low-energy effective gravitational theories, such as the Einstein-Hilbert
theory \refb{action-Einstein} or SUGRA theories \refb{action-strings}.
Therefore, production of NPGOs which are created in hard processes with CM
energy $E_{CM}\gg M_{\rm Pl}$ can be described by semiclassical gravitational
theories, just as gravitational collapse is described by general
relativity.~\footnote{Formation of NPGOs by inelastic super-Planckian
collisions can be thought as asymmetric gravitational collapse.}

\begin{table}[h]
\tbl{Regimes and phenomenology of gravity as a function of the energy. As is
explained in the text, at super-Planckian energies we expect formation of NPGOs
that are described by semiclassical low-energy gravitational theories.}
{\begin{tabular}{@{}lcccc@{}} \toprule 
Energy&Gravity regime&QG effects& Theory&Phenomenology\\ \colrule
$E_{CM}\laq\, M_\star$ & perturbative & no & SM + KK theory& KK states \\
$E_{CM}\sim M_\star$ & non-perturbative & yes & Strings/M-theory &
String balls/Branes? \\
$E_{CM}\gaq\, M_\star$ & non-perturbative & no & Einstein/Sugra & BHs/p-Branes \\
\botrule
\end{tabular}}
\end{table}
\subsection{Super-Planckian BHs}
BH formation is a generic non-perturbative process of any gravitational
theory. A BH smaller than the size of all LEDs ``sees'' a $D$-dimensional
isotropic spacetime, thus is spherically symmetric. A non-rotating
spherically symmetric BH is described by the $(n+4)$-dimensional Schwarzschild
solution:
\be
ds^2=-R(r)dt^2+R(r)^{-1}dr^2+r^2d\Omega^2_{n+2}\,,
\label{Schwarzschild}
\ee
where
\be
R(r)=1-\left({r_{s}\over r}\right)^{n+1}\,.
\label{R-Schw}
\ee
The Schwarschild radius $r_s$ is related to the mass $M_{\rm BH}$ by the
relation
\be
r_{s} = {1\over\sqrt{\pi}M_{\star}}\gamma(n)\,
\left({M_{\rm BH}\over M_{\star}}\right)^{1\over n+1}\,,
\label{rs-Mbh}
\ee
where 
\be
\displaystyle
\gamma(n)=\left[{8\,\Gamma \left({n+3\over 2}\right)\over
(2+n)}\right]^{1\over n+1} \,.
\label{gamma-Schw}
\ee
BHs with mass $M_{\rm BH}\sim M_\star$ have Schwarzschild radius $r_s\sim
M_\star^{-1}$. In symmetric compactification models the size of extra
dimensions is much larger than $M_\star^{-1}$. Therefore, the spherical
approximation is justified. The latter breaks down for asymmetric
compactifications with some of the extra dimensions of order of the fundamental
Planck scale. In this case the geometry of non-perturbative objects is that of
black strings and branes. A BH with angular momentum $J$ is described by the
Kerr solution. Equation \refb{rs-Mbh} is substituted by
\be
r_{s} = {1\over\sqrt{\pi}M_{\star}}\gamma(n)\,
\left({M_{\rm BH}\over M_{\star}}\right)^{1\over n+1}
\left[1+{(n+2)^2J^2\over 4r_s^2 M_{\rm BH}}\right]^{-{1\over n+1}}\,.
\label{rs-Mbh-kerr}
\ee
The radius of a spinning BH is smaller than the radius of a Schwarzschild BH
of equal mass. 

The $D$-dimensional Schwarzschild BH has different mechanical and
thermodynamical properties from its four-dimensional analogue. Argyres,
Dimopoulos and March-Russell~\cite{Argyres:1998qn} have discussed the
properties of spherically symmetric black holes in higher-dimensional
spacetimes. The Hawking temperature and the entropy of a $(n+4)$-dimensional BH
are
\be
T_H={n+1\over 4\pi r_s}\,,\qquad S={(n+1)M_{\rm BH}\over (n+2)T_H}\,,
\ee
respectively. Neglecting factors of order $O(1)$, the lifetime of a BH with
mass $M_{\rm BH}$ is
\be
\tau\sim {1\over M_\star}\left({M_{\rm BH}\over M_\star}\right)^{n+3\over n+1}\,.
\label{bh-lifetime}
\ee
Higher-dimensional BHs are colder, longer-lived and with a greater radius than
four-dimensional BHs of equal mass. Thus BH production is easier in higher
dimensions.

At super-Planckian energies we expect formation of either Schwarzschild or Kerr
BHs. The classical description is valid if the entropy of the process is
sufficiently large, i.e., the fluctuations of the number of (micro) canonical
degrees of freedom are small. BHs with mass equal to a few Planck masses
usually satisfy this condition. For example, in $D=10$ dimensions the entropy
of a Schwarzschild BH with mass equal to 5 (10) times the fundamental scale is
$S\sim 8$ (17). The lifetime of these BHs is larger than the inverse of their
mass. From Eq.~\refb{bh-lifetime} it follows that $\tau M_{\rm BH}\sim (M_{\rm
BH}/M_\star)^{2(n+2)/(n+1)}$. For a ten-dimensional BH with mass $M_{\rm
BH}\sim 5~(10)~M_\star$ we have $\tau M_{\rm BH}\sim 40~(190)\gg 1$. The BHs
formed in super-Planckian collisions can be thought as long-lived intermediate
states, i.e., resonances.
\subsection{BH Production}
The physics of BH formation in hard collisions has been first described in
detail by Banks and Fischler~\cite{Banks:1999gd} and later by Giddings and
Thomas in Ref.~\cite{Giddings:2001bu}. Super-Planckian particle collision
processes are dominated by BH formation in the $s$-channel. The initial state
is described by two incoming Aichelburg-Sexl shock waves with impact parameter
$b$. If the impact parameter is smaller than the Schwarzschild radius
associated with the CM energy of the incident particles, an event horizon
forms. At super-Planckian energies the process of BH formation is
semiclassical. Therefore, the cross-section is approximated by the geometrical
cross section of an absorptive black disk with radius $r_s$:
\be
\sigma_{ij\to BH}(s;n)=F(s)\pi r_{s}^2\,,
\label{sigma}
\ee
where $\sqrt{s}$ is the CM energy of the colliding particles and $F(s)$ is a
dimensionless form factor of order one. Using Eqs.~\refb{rs-Mbh} and
\refb{gamma-Schw} the cross section for a non-rotating BH is
\be
\displaystyle
\sigma_{ij\to BH}(s;n) =F(s){1\over s_{\star}}\gamma(n)^2\,
\left({s\over s_{\star}}\right)^{1\over n+1}\,,
\label{cross-BH}
\ee
where $s_{\star}=M_{\star}^2$. The form factor $F(s)$ reflects the theoretical
uncertainties in the dynamics of the process, such as the amount of initial CM
energy that goes into the BH, the distribution of BH masses as function of the
energy, and corrections to the geometrical black disk cross section. $F(s)$ is
usually chosen equal to one. Possible corrections to the cross section
\refb{cross-BH} have been described by Anchordoqui {et al.} in
Ref.~\cite{Anchordoqui:2001cg}. We summarize them here:
\begin{itemize}
\item[a)] {\bf Mass ejection corrections.} Numerical simulations suggest that,
at least in four-dimensions and in head-on collisions, the mass of the BH is
less than the CM energy of the incoming particles
\cite{D'Eath:hb,D'Eath:hd,D'Eath:qu}, i.e., the scattering is not completely
inelastic. This result seems to suggest that $F(s)\laq 1$.~\footnote{See also
Ref.~\cite{Kohlprath:2002yh}.} 
\item[b)] {\bf Angular momentum corrections.} Equation \refb{rs-Mbh-kerr}
suggests that BHs with nonvanishing angular momentum have smaller cross
sections. The correction factor has been estimated by Anchordoqui {\it et
al.}~\cite{Anchordoqui:2001cg}. They found that angular momentum corrections
lead typically to a reduction of the cross-section of about 40\%.~\footnote{See
also Ref.~\cite{Park:2001xc}.}
\item[c)] {\bf Sub-relativistic limit.} A naive argument based on the
non-relativistic limit of a two-BH scattering suggests that the geometrical
cross section can be enhanced by a factor $\laq\, 400$\%.
\item[d)] {\bf Gravitational infall corrections.}
Solodukhin~\cite{Solodukhin:2002ui} used the classical cross section for
photon capture of a BH to estimate the cross section of BH formation. With this
definition, the cross section is enhanced by a factor ranging from 400\%
($n=1$) to 87\% ($n=7$).
\item[e)] {\bf Voloshin suppression.} The most controversial correction to the
cross section \refb{cross-BH} has been proposed by
Voloshin~\cite{Voloshin:2001vs,Voloshin:2001fe}. Voloshin's claim is that BH
creation must be described by an instanton. Thus the black disk cross section
\refb{cross-BH} has to be multiplied by a suppression factor proportional to
$e^{-S_E}$, where $S_E$ is the BH Euclidean action. Although the controversy
has not yet been completely solved, numerical simulations in head-on
collisions~\footnote{See a).} seem to contradict Voloshin's result.
\end{itemize}
In the following sections we will set $F(s)=1$. However, as a reminder of a
possible uncertainty of order $O(1)$, we will also replace the equal sign with
$\sim$.

\begin{figure}
\centerline{\psfig{file=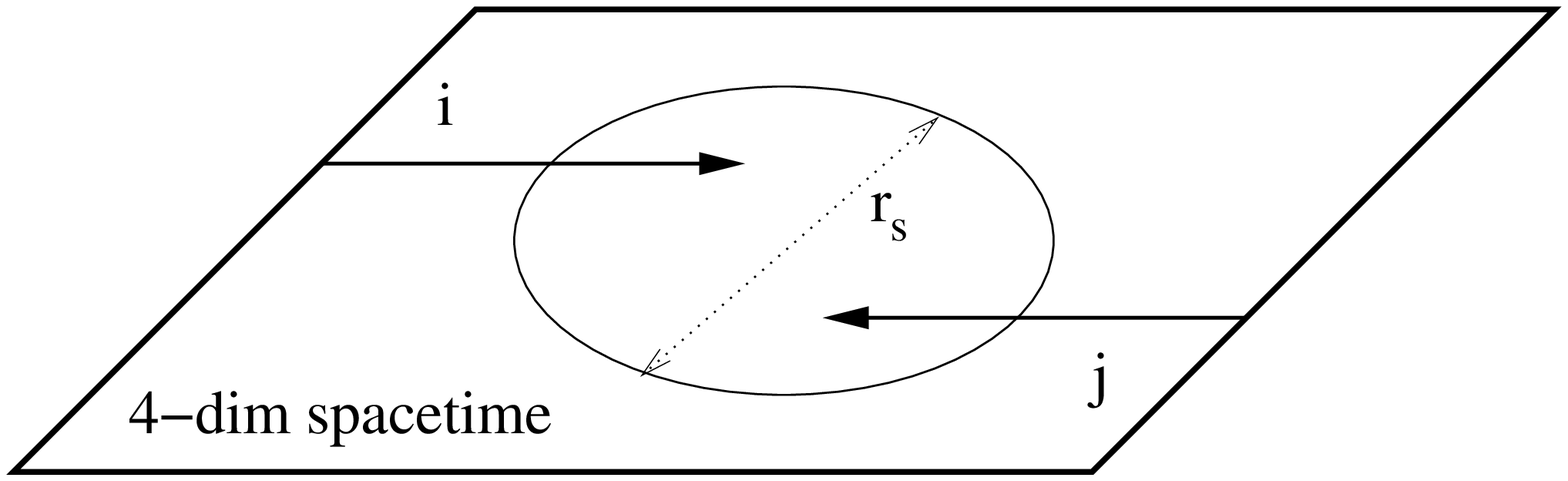,height=16mm}\qquad
\vbox to 16mm{\vfill\hbox{$\to$}\vfill}\qquad
\psfig{file=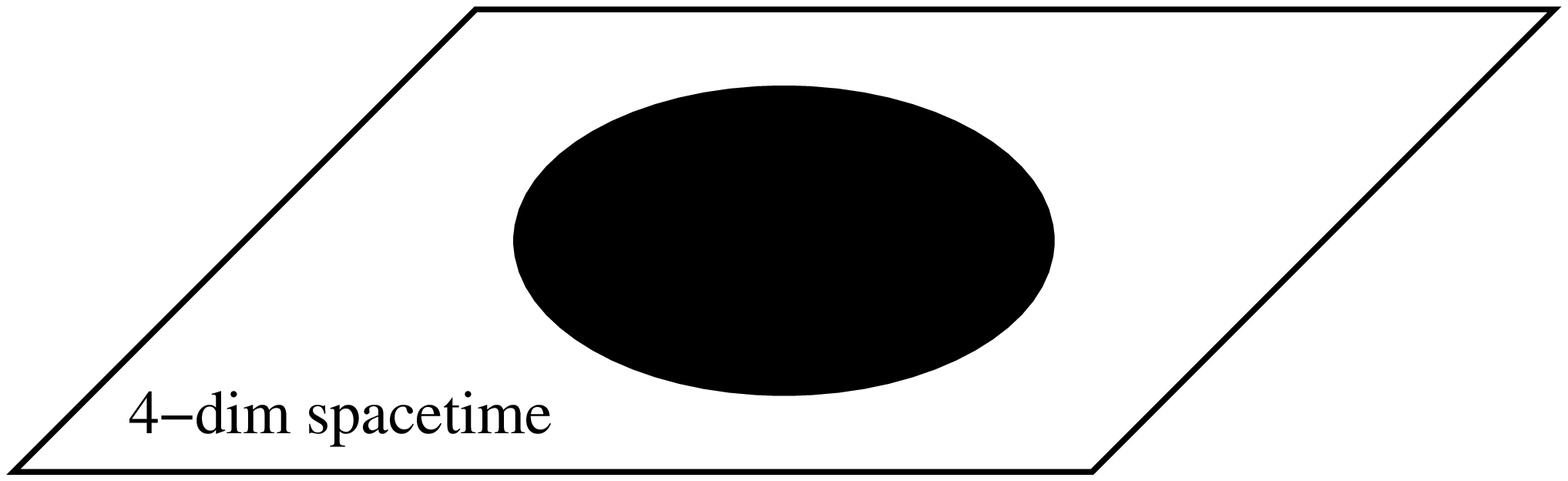,height=16mm}}
\vspace*{8pt}
\caption{Schematic illustration of BH formation by super-Planckian scattering
of two incident particles $i$ and $j$ with impact parameter $b$. The event
horizon forms before the particles come in causal contact.}
\end{figure}

The Schwarzschild radius of a BH with mass $M_{\rm BH}\sim M_\star$ is of order of
the fundamental Planck length. Therefore, in a hadronic collision, the cross
section \refb{cross-BH} must be interpreted at the parton level.
For a proton-proton ($pp$) and a neutrino-proton ($\nu p$) collision the total
cross sections are
\be
\sigma_{pp\to BH}(x_m;n)\sim\sum_{ij}\int_{x_m}^1dx\,\int_x^1 {dy\over
y}\,f_i(y,Q)f_j(x/y,Q)\,\sigma_{ij\to BH}(xs;n)\,,
\label{cross-pp}
\ee
and
\be
\sigma_{\nu p\to BH}(x_m;n)\sim\sum_{i}\int_{x_m}^1dx\,
f_i(x,Q)\,\sigma_{ij\to BH}(xs;n)\,,
\label{cross-nup}
\ee
respectively. Here, $\sqrt{sx_m}=M_{\rm BH,min}\sim$ few $M_\star$ is the minimal BH
mass for which the semiclassical cross section \refb{cross-BH} is valid, $f_i$
are the Parton Distribution Functions (PDFs), and $Q$ is the momentum
transfer.~\footnote{In numerical calculations throughout the paper we will use
the CTEQ6 PDFs \cite{CTEQ6}.} The sum over partons in Eq.~\refb{cross-pp} and
Eq.~\refb{cross-nup} leads to a big enhancement of the BH cross section w.r.t.\
cross section of perturbative SM processes.
\subsection{BH Decay and Experimental Signatures}
A lot of efforts have been devoted to the study of the experimental signatures
of BH production. After its formation, a BH evaporates semiclassically in a
time $t\sim 10^{-25}$ sec. by emitting Hawking
radiation~\cite{Hawking:sw}.~\footnote{The lifetime of a BH in the
microcanonical picture is longer than in the canonical picture. In particular,
in the RS model the evaporation process of BHs could be frozen at the
fundamental Planck scale (see Ref.~\cite{Casadio:2001wh}).} The decay phase is
divided in three stages. In the first stage the BH sheds the hair associated
with multipole momenta by emitting gauge radiation (SM fields on the brane and
gravitons in the bulk). In the second stage the BH loses angular momentum,
before evaporating by emission of thermal Hawking radiation with temperature
$T_H$ (last stage). The Hawking evaporation ends when the mass of the BH approaches
$\sim M_\star$. At this point the semiclassical description breaks down (see
Table 4) and the BH is believed to decay completely by emitting a few quanta
with energy of order of $M_\star$. 

\begin{figure}
\centerline{\psfig{file=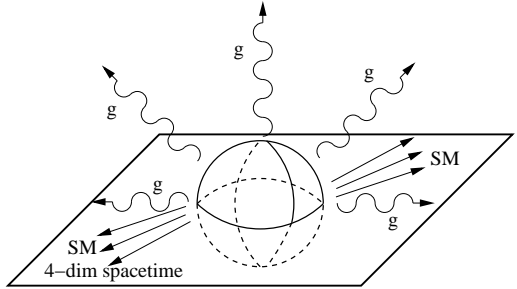,width=7cm}}
\vspace*{8pt}
\caption{Schwarzschild stage of BH evaporation. The BH emits brane and bulk
modes. The former are SM fields that can be observed. The gravitons $g$ are
emitted in the bulk and cannot be observed.}
\end{figure}

Emparan {\it et al.}~\cite{Emparan:2000rs} have found that BHs decay by
emitting mainly on the brane. This can be qualitatively understood as follows.
Since the wavelength of the Hawking thermal spectrum is much larger than the
size of the BH, the latter evaporates in the $s$-channel. Therefore, BHs decay
equally in all modes. Since the SM fields only propagate on the brane, the
number of states emitted on the brane is greater than the number of states
emitted in the bulk. Excluding the Higgs boson(s), the ratio of the degrees of
freedom of SM gauge bosons, quarks and leptons on a brane with infinitesimal
thickness is 29:72:18. This leads to a final hadronic to leptonic activity
roughly 5:1 and a ratio of hadronic to photonic activity of about 100:1. The
experimental signatures of BH decay have been summarized in
Ref.~\cite{Giddings:2001bu}. The most important are:
\begin{itemize}
\item[a)] Hadronic to leptonic activity of roughly 5:1;~\footnote{Recently, Han
{\it et al.} \cite{Han:2002yy} have investigated BH evaporation on a FB
(see Sect.~2.3.2). They find that the hadronic to leptonic to photonic activity
is 113:8:1.}
\item[b)] High multiplicity; 
\item[c)] High sphericity;~\footnote{This is valid if the BH is at rest in the
CM frame, i.e., for the completely inelastic collision $ij\to$ BH. If SM
particles are present in the final state, i.e., the process is $ij\to {\rm
BH}+k$, the BH is boosted and the decay is not spherical~\cite{Cheung:2002aq}.}
\item[d)] Visible transverse energy of order $\sim 30$\% of the total energy;
\item[e)] Emission of a few hard visible quanta at the end of the evaporation
phase;
\item[f)] Suppression of hard perturbative scattering processes.
\end{itemize}

The events that can potentially lead to BH production are essentially
high-energy scattering in particle colliders and UHECR. The next generation of
particle colliders are expected to reach energies above 10 TeV. LHC~\cite{LHC}
(in construction at CERN) and VLHC~\cite{VLHC} are planned to reach a CM energy
of 14 and 100 TeV with a luminosity ${\cal F}$ of $\sim 300$ fb$^{-1}$
yr$^{-1}$, respectively. Therefore, if the fundamental Planck scale is of order
of few TeV, LHC and VLHC would copiously produce BHs. A number of authors have
studied the production rates of BH at LHC and other particle
colliders~\cite{Rizzo:2001dk,Rizzo:2002kb,Cheung:2001ue}.
\begin{figure}
\centerline{\psfig{file=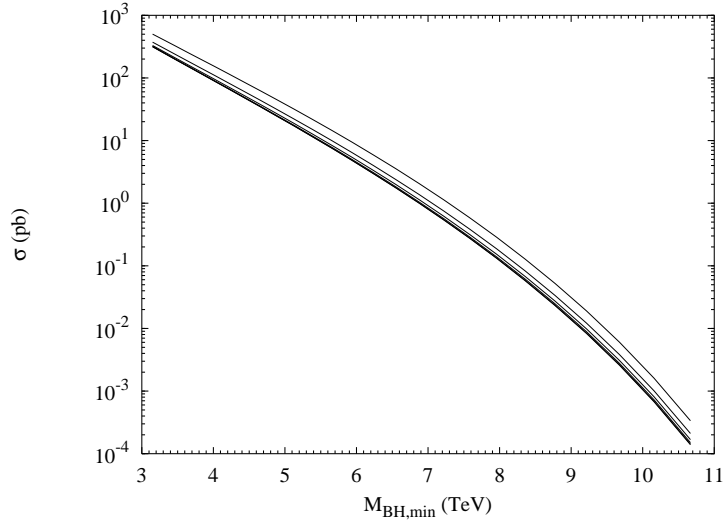,width=7cm, angle=-90}}
\vspace*{8pt}
\caption{Total cross sections (pb) for BH formation ($n=2\dots 7$ from above) by
proton-proton scattering at LHC (CM energy = 14 TeV). We have assumed
$M_\star=1$ TeV and a threshold for BH formation $M_{\rm BH,min}=3M_\star$.}
\end{figure}
In Fig.~4 we present the total cross sections for events at LHC. For this
particular choice of parameters the cross section is in the range $10^{-3}$ pb
(high thresholds) -- $10^2$ pb (low thresholds). With a luminosity of $\sim
3\cdot 10^5$ pb$^{-1}$ yr$^{-1}$, LHC might create BHs at a rate of one event
per second! Rizzo has calculated the cross section and the production
rate of BH formation at LHC when the Voloshin suppression is active. Though
greatly reduced, the cross section is still sufficiently large to allow
observation of BH formation at LHC~\cite{Rizzo:2001dk,Rizzo:2002kb}. For
instance, for a mimimum BH mass of 5 TeV and $M_\star=1$ TeV, the integrated
cross section \refb{cross-pp} is in the .1 -- 1 pb range. Finally, other
interesting possible signatures of BH formation at particle collider have been
discussed by Uehara~\cite{Uehara:2002cj,Uehara:2002gv} and Anchordoqui and
Goldberg  ~\cite{Anchordoqui:2002cp}.

BH production by cosmic rays has also been recently investigated by a number of
authors~\cite{Feng:2001ib,Ringwald:2001vk,Anchordoqui:2001cg,Kowalski:2002gb,Alvarez-Muniz:2002ga,Anchordoqui:2001ei}.
Cosmogenic neutrinos~\cite{Engel:2001hd} with energies above the
Greisen-Zatsepin-Kuzmin (GZK) cutoff~\cite{Greisen:1966jv,Zatsepin:1966jv} are
expected to create BHs in the terrestrial atmosphere. The thermal decay of the
BHs produces air showers which could be observed.  The cross sections of these
events are two or more orders of magnitude larger than the cross sections of SM
processes (see Fig.~5). Therefore, BHs are created uniformly at all atmospheric
depths with the most promising signal given by quasi-horizontal showers which
maximize the likelihood of interaction. This allows BH events to be
distinguished from other SM events.

\begin{figure}
\centerline{\psfig{file=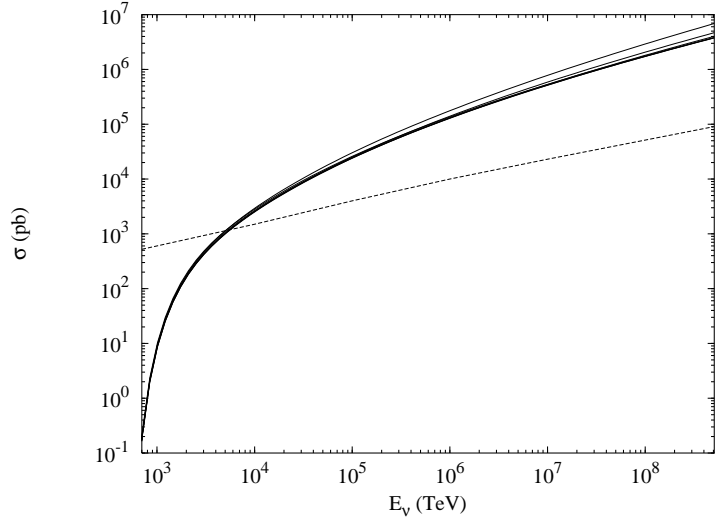,width=7cm, angle=-90}}
\vspace*{8pt}
\caption{Total cross sections (pb) of BH production by UHECR (neutrinos of
energy $E_\nu$, $n=2\dots 7$ from above). The dashed curve is for the SM
process. We have assumed $M_\star=1$ TeV $=M_{\rm BH,min}$.}
\end{figure}

The current non-observation of BH events at particle colliders and of BH-induced
air showers in cosmic ray experiments sets lower bounds on the fundamental
Planck scale. Bleicher {\it et al.}~\cite{Bleicher:2001kh,Hofmann:xd} find
$M_\star\gaq 1.4$ TeV from Tevatron data (CM energy = 1.8 TeV). The lower
bounds on the fundamental Planck scale derived from UHECR experiments are given
in Refs.~\cite{Feng:2001ib,Ringwald:2001vk,Anchordoqui:2001cg}. Data from the
Akeno Giant Air Shower Array~\cite{AGASA} (AGASA) set a lower bound of
$M_\star= .65$ -- .75 TeV for $n=4$ and $M_\star=.55$ -- .62 TeV for $n=7$.
Future UHECR experiments such as Auger~\cite{Auger} will either set more
stringent limits on $M_\star$ or detect several BH events per year.
\section{Brane Factories}
Super-Planckian particle collisions may create extended objects other than BHs.
The simplest extended object in $D$ dimensions is described by the metric
\be
ds^2=g_{ab}(x)dx^a dx^b=
\gamma_{\mu\nu}(y)dy^\mu dy^\nu+f(y)\delta_{mn}dz^m dz^n\,,
\label{brane.generic}
\ee
where $\mu,\nu=0,\dots D-p-1$, $m,n=D-p,\dots D-1$, and $\gamma_{\mu\nu}$ is a
pseudo-Rieman\-nian metric in $D-p$ dimensions. Equation \refb{brane.generic}
includes as special cases the $p$-brane metrics mentioned in the Introduction.
The $\hbox{(Poincar\'e)}_{p+1}\times {\rm SO}(N-p-1)$ symmetry is obtained by
setting $\gamma_{00}=-f(y)\equiv -f(r)$, $\gamma_{0i}=0$, and
$\gamma_{ij}=g(r)\delta_{ij}$, where $r=(y_iy^i)^{1/2}$ and $i,j=1,\dots
D-p-1$. In ``Schwarzschild-like'' coordinates the $p$-brane metric is
\be
ds^2=A(r)(-dt^2+dz_i^2)+B(r)dr^2+r^2C(r)d\Omega^2_{q}\,.
\label{pbrane}
\ee
$p$-branes arise as solutions of low-energy effective gravity and have been
widely studied in the literature. In the next two sections we will briefly
review brane solutions in Einstein-Maxwell gravity and low-energy effective
STs.
\subsection{Einsteinian Branes}
The $p$-brane solutions of Einstein-Maxwell-dilaton gravity have been discussed
by Gregory~\cite{Gregory:1995qh}. The general uncharged non-spinning
Einsteinian brane solution of $(n+4)$-dimensional Einstein gravity
is~\cite{Cavaglia:1997hc}  
\be
ds^2=-R^{\kappa\delta}dt^2+R^{{\kappa\over p+q}}dz_i^2+R^{{1\over q-1}
\left[1-\kappa\left(\delta+{p\over p+q}\right)\right]}\left(R^{-1}dr^2+
r^2d\Omega^2_q\right)\,,
\label{brane.einstein}
\ee
where 
\be
R=1-\left({r_p\over r}\right)^{q-1}\,,
\label{Rbrane}
\ee
and $\kappa$ and $\delta$ are two parameters related by
\be
|\kappa|=\left[\delta\left(\delta+{2p\over q(p+q)}\right)+{p\over
q(p+q)}\left(1-{1\over p+q}\right)\right]^{-1/2}\,.
\ee
The line element is singular on the hypersurface $r=r_p$. The nature of the
singularity depends on the value of the modulus $\delta$. Setting
\be
\tilde R(\tilde r)\equiv\left[1+\left({r_p\over\tilde r}\right)^{q-1}\right]
=\left[1-\left({r_p\over r}
\right)^{q-1}\right]^{-1}\equiv R^{-1}(r)\,,
\label{Rtilde-transf}
\ee
the metric is cast in the form
\be
ds^2=-\tilde R^{-\kappa\delta}dt^2+\tilde R^{-{\kappa\over p+q}}dz_i^2+
\tilde R^{{1\over q-1}
\left[1+\kappa\left(\delta+{p\over p+q}\right)\right]}\left(\tilde R^{-1}
d\tilde r^2+
\tilde r^2d\Omega^{(q)}\right)\,.
\label{brane.einstein2}
\ee
The metric \refb{brane.einstein2} covers only the exterior part of the
$p$-brane spacetime. Equation \refb{brane.einstein2} is obtained from
Eq.~\refb{brane.einstein} with the substitution $\kappa\to -\kappa$ and
$r_p^{q-1}\to -r_p^{q-1}$. In the following we will choose $\kappa\ge 0$
without loss of generality. The (boosted) $p$-brane solution is obtained by
choosing $\delta=(p+q)^{-1}$. The line element is
\be
ds^2=R^{{\Delta\over p+1}}(-dt^2+dz_i^2)+R^{{2-q-\Delta\over
q-1}}dr^2+r^2R^{{1-\Delta\over q-1}}d\Omega^2_{q}\,,
\label{brane.einstein3}
\ee
where
\be
\Delta=\sqrt{{q(p+1)\over p+q}}\,.
\label{Delta}
\ee
The spherically symmetric solution is recovered for $p=0$; in this case we have
$\Delta=1$, and Eq.\ \refb{brane.einstein3} reduces to the $(n+4)$-dimensional
Schwarzschild BH. Equation \refb{brane.einstein3} describes an asymptotically
flat spacetime. When $p=0$ (BH) $r=r_{p}\equiv r_s$ defines the Schwarzschild
horizon. For $p\not=0$ the metric \refb{brane.einstein3} possesses a
non-conical \cite{chk:2002} naked singularity at $r=r_{p}$ which is the
higher-dimensional analogue of a cosmic string singularity. $r_{p}$ can be
interpreted as the ``physical radius'' of the $p$-brane. The interpretation of
the curvature singularity has been discussed in Ref.\ \cite{Gregory:1995qh}.
The metric \refb{brane.einstein3} is interpreted as {\em vacuum} exterior
solution to the $p$-brane, with the curvature singularity being smoothed out by
the core of the $p$-brane. 

The previous solutions can be generalized to include electromagnetic charge and
dilaton field \cite{Gregory:1995qh}. In the Einstein frame
\refb{action-strings} the general black brane solution
is \cite{Stelle:1996tz} ($a=1/2$)
\begin{eqnarray}
ds^2&=&-R_+R_-^{{2\over p+1}\left(1+{qb^2\over
2\Delta^2(q-1)}\right)^{-1}-1}dt^2+
R_-^{{2\over p+1}\left(1+{qb^2\over
2\Delta^2(q-1)}\right)^{-1}}
dz_i^2+\nonumber\\
&+&R_+^{-1}R_-^{{2\over q-1}\left(1+{2\Delta^2(q-1)\over qb^2
}\right)^{-1}-1}dr^2+
r^2 R_-^{{2\over q-1}\left(1+{2\Delta^2(q-1)\over qb^2
}\right)^{-1}}d\Omega^2_{q}\,,\\
\phi&=&\phi_0\mp{2\over b}\left[1+{
2\Delta^2(q-1)\over qb^2}\right]^{-1}\ln R_-,
\label{black-brane}
\end{eqnarray}
where $\phi$ is the dilaton field and
\be 
R_\pm=1-\left({r_\pm\over r}\right)^{q-1}\,.
\label{Rbrane-charged}
\ee
The boosted magnetically charged solution with EM field $F_{[n]}=Q\epsilon_q$
has been given by Gregory in Ref.~\cite{Gregory:1995qh}.
\begin{eqnarray}
ds^2&=&e^{{2(q-1)\over b(p+q)}\phi}R_+^{{\Delta\over p+1}}
R_-^{-{\Delta\over p+1}}(-dt^2+dz_i^2)+
e^{-{2(p+1)\over b(p+q)}\phi}R_+^{{2-q-\Delta\over
q-1}} R_-^{{2-q+\Delta\over
q-1}} dr^2+\nonumber\\
&&\hbox to 51truemm{\hfill}
+e^{-{2(p+1)\over b(p+q)}\phi}r^2R_+^{{1-\Delta\over q-1}}
R_-^{{1+\Delta\over q-1}}d\Omega^2_{q}\,,
\label{brane-charged}
\end{eqnarray}
\be
\phi=\phi_0-{2\over b}\left[1+{
2\Delta^2(q-1)\over qb^2}\right]^{-1}\ln\left[1-{1-R_+^\Delta R_-^{-\Delta}\over
\Delta(1-r_+^{q-1}r_-^{1-q})}\right],
\label{dil-charged}
\ee
where 
\be
Q^2={4(q-1)^2\over b^2}\left[1+{2\Delta^2(q-1)\over qb^2}\right]^{-1}
r_-^{2(q-1)}[1-\Delta(1-r_+^{q-1} r_-^{1-q})]
\ee
The electrically charged solution is obtained by a duality
transformation~\cite{Gregory:1995qh}. A $p$-brane is always dual to a
$(D-p-4)$-brane. If the spacetime dimension is even, self-dual branes with
$p=D/2-2$ and $q=D/2$ exist. Self-dual branes are discussed in
Ref.~\cite{Gregory:1995qh}. The extremal limit of Eqs.~\refb{black-brane} and
\refb{brane-charged} is obtained by setting $r_+=r_-$.~\footnote{Note that the
extremal limit of the generic black brane solution is a boosted brane.} The
metric and the dilaton are
\begin{eqnarray}
ds^2&=&R^{{2\over p+1}\left(1+{qb^2\over
2\Delta^2(q-1)}\right)^{-1}}(-dt^2+dz_i^2)+
R^{{2\over q-1}\left[1-q+\left(1+{2\Delta^2(q-1)\over qb^2
}\right)^{-1}\right]}dr^2+\nonumber\\
&&\hbox to 45truemm{\hfill}+r^2 R^{{2\over q-1}\left(1+{2\Delta^2(q-1)\over qb^2
}\right)^{-1}}d\Omega^2_{q}\,,\\
\phi&=&\phi_0\mp{2\over b}\left[1+{
2\Delta^2(q-1)\over qb^2}\right]^{-1}\ln R\,.
\label{extremal}
\end{eqnarray}
If we set $b\to 0$ in Eq.~\refb{black-brane} the dilaton field decouples. The general Einstein-Maxwell solution is
\be
ds^2=-R_+R_-^{1-p\over 1+p}dt^2+
R_-^{2\over 1+p}dz_i^2+(R_+R_-)^{-1}dr^2+r^2 d\Omega^2_{q}
\label{EM-brane}
\ee
The extremal Einstein-Maxwell brane is obtained by setting $R_+=R_-\equiv R$:
\be
ds^2=R^{2\over 1+p}(-dt^2+dz_i^2)+R^{-2}dr^2+r^2 d\Omega^2_{q}\,.
\label{EM-brane-extremal}
\ee
Setting $p=0$ in Eqs.~\refb{EM-brane} and \refb{EM-brane-extremal} we recover the Reissner-Nordstr\"om solution.
\subsection{String and Supergravity Branes}
The general solution of the previous section includes branes of SUGRA theories.
Let us see a few examples. Choosing
\be
b=-\sqrt{2\over p+q}(q-1)
\ee
in Eq.~\refb{brane-charged}, and passing to the string frame
\be
g^s_{ab}=\exp\left[\sqrt{2\over p+q}\phi\right]g_{ab}\,,
\ee
Eq.~\refb{brane-charged} and \refb{dil-charged} read
\begin{eqnarray}
ds^2&=&R_+^{{\Delta\over p+1}}
R_-^{-{\Delta\over p+1}}(-dt^2+dz_i^2)+e^{{\sqrt{2(p+q)}\over q-1}\phi}
R_+^{1-\Delta\over q-1}R_-^{1+\Delta\over q-1}\cdot\nonumber\\
&&\hbox to 5truecm{\hfill}\cdot\left[(R_+ R_-)^{-1}dr^2+
r^2d\Omega^2_{q}\right]\,,
\label{brane-special}\\
\phi&=&\phi_0+\sqrt{2\over p+q}\ln\left[1-{1-R_+^\Delta R_-^{-\Delta}\over
\Delta(1-r_+^{q-1}r_-^{1-q})}\right]\,.
\end{eqnarray}
The non-supersymmetric five brane in ten dimensions is obtained by setting $p=5$
and $q=3$:
\be
ds^2=\left({R_+\over R_-}\right)^{1/4}
(-dt^2+dz_i^2)+e^{2\phi}
R_+^{-1/4}R_-^{5/4}
\left[(R_+ R_-)^{-1}dr^2+r^2d\Omega^2_{3}\right]\,.
\label{brane-non-susy}
\ee
The six-brane of ten-dimensional heterotic ST is obtained by
choosing $p=6$ and $q=2$:
\be
ds^2=\left({R_+\over R_-}\right)^{1\over 2\sqrt{7}}
(-dt^2+dz_i^2)+e^{4\phi}\left({R_+\over R_-}\right)^{-\sqrt{7}\over 2}
\left[dr^2+R_+ R_-r^2d\Omega^2_{2}\right]\,.
\label{brane-het-magn}
\ee
The electrically charged solution dual to the six-brane \refb{brane-het-magn}
is the zero-brane
\be
ds^2=-R_+R_-dt^2+R_+^{-1}R_-^{-5/7}dr^2+r^2R_-^{2/7}
d\Omega^2_{8}\,.
\label{brane-het-el}
\ee
Setting $p=5$ and $q=4$ in Eq.~\refb{EM-brane-extremal} we obtain the
(extremal) solitonic/magnetic five-brane of 11-dimensional
SUGRA~\cite{Stelle:1996tz}:
\be
ds^2=R^{1/3}(-dt^2+dz_i^2)+R^{-2}dr^2+r^2d\Omega^2_{4}\,.
\label{brane-sugra-magn}
\ee
The corresponding dual electric two-brane is \cite{Stelle:1996tz}
\be
ds^2=R^{2/3}(-dt^2+dz_i^2)+R^{-2}dr^2+r^2d\Omega^2_{7}\,.
\label{brane-sugra-el}
\ee
The magnetic five-brane \refb{brane-sugra-magn} and the electric two-brane
\refb{brane-sugra-el} saturate the Bogomol'ny bound \cite{Duff:1996hp} (BPS
solutions). Thus they leave some portion of supersymmetry unbroken. It can
be shown\cite{Stelle:1996tz} that Eqs.~\refb{brane-sugra-magn} and
\refb{brane-sugra-el} preserve half of the rigid $D=11$ supersymmetry.
\subsection{$p$-brane Mass and Radius}
The parameters $r_p$ and $r_\pm$ are related to the ADM \cite{MTW} mass $M_p$
and to the electric/magnetic charge $Q$ of the brane. The general formula for
the ADM mass of a $p$-brane has been derived by Lu \cite{Lu:1993vt}. Given a
generic metric of the form
\be
ds^2=-A(r)dt^2+D(r)dz_i^2+B(r)dr^2+r^2C(r)d\Omega^2_{q}\,,
\label{pbrane-gen}
\ee
the ADM mass of the brane is:
\be
M_p=M_\star^{q}v_p{\pi^{q-1\over 2}\over
8\Gamma\left({q+1\over 2}\right)}
\left\{r^q\left[q\partial_r C(r)+p\partial_r
D(r)\right]+qr^{q-1}\left[B(r)-C(r)\right]\right\}_{r\to\infty}\,,
\ee
where $v_p$ is the $p$-dimensional volume of the brane in fundamental Planck
units. For instance, the ADM mass of the neutral Einsteinian $p$-brane
\refb{brane.einstein3} in $n+4$ dimensions is
\be
M_{p}=M_\star^{n+2-p}v_p\left({\sqrt{\pi}r_p\over\gamma(n,p)}\right)^{n+1-p}\,,
\ee
where 
\be
\gamma(n,p)=\left[8\,\Gamma \left({n+3-p\over 2}\right)
\sqrt{p+1\over (n+2)(n+2-p)}\right]^{1\over
n+1-p} \,.
\label{gamma}
\ee
Inverting Eq.~\refb{gamma} we obtain the $p$-brane radius
\be
\displaystyle
r_{p} = {1\over\sqrt{\pi}M_{\star}} \gamma(n,p)\,v_p^{-{w\over n+1}}
\left({M_{p}\over M_{\star}}\right)^{w\over n+1}\,.
\label{rp}
\ee
where $w=[1-p/(n+1)]^{-1}\ge 1$. For $p=0$ (0-brane) $w=1$ and we recover
Eqs.~\refb{rs-Mbh} and \refb{gamma-Schw}. 
\subsection{Brane Formation in TeV Gravity}
In analogy to the BH case, scattering of two partons with impact parameter
$b\,\laq\, r_{p}$ produces a $p$-brane which is described by a suitable
localized energy field configuration and whose exterior geometry has metric
\refb{pbrane}. Assuming that the collision is completely inelastic, the cross
section for the process depends on the brane tension. The geometrical cross
section corresponding to the black absorptive disk of radius
$r_p$~\cite{Ahn:2002mj} is:
\be
\sigma_{ij\to br}=F(s)\pi r_{p}^2\,.
\label{sigma-brane}
\ee
Setting $F(s)=1$~\footnote{See Sect.~3.2.}, the cross section
\refb{sigma-brane} is given by  
\be
\sigma_{ij\to br}(s;p,n,v_p)\sim{1\over s_\star}G(n,p)\,v_p^{-{2\over n-p+1}}
\left({s\over s_\star}\right)^{1\over n-p+1}\,,
\label{cross-brane}
\ee
where $s=E_{ij}^2$ is the square of the CM energy of the two
scattering partons, and $s_\star=M_\star^2$. The function $G(n,p)$
depends on the model considered. For instance,
\be
\displaystyle
G_{\rm Ein}(n,p)=\gamma(n,p)^2=\left[64(p+1)\,\Gamma \left[(n+3-p)/2\right]^2
\over (2+n)(n-p+2)\right]^{1\over n-p+1} \,,
\label{G-Einstein}
\ee
for the Einsteinian brane and 
\be
G_{\rm el}(n,p)=2\,,\qquad G_{\rm mg}(n,p)=(2\sqrt{\pi})^{2/3}\,,
\ee
for the electric and magnetic SUGRA branes, respectively. 

\begin{figure}
\centerline{\vbox to 64mm{\vfill\null\vskip -20pt
\hbox{\psfig{file=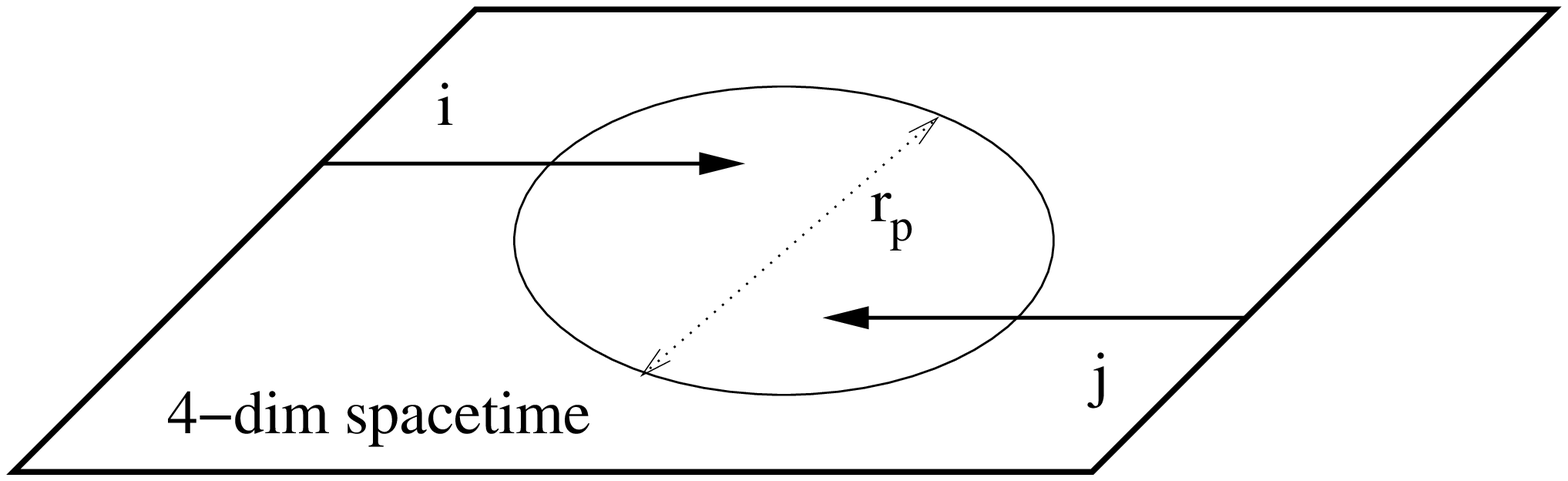,height=16mm}}\vfill}
\qquad\vbox to 64mm{\vfill\hbox{$\to$}\vfill}\qquad
\vbox to 64mm{\vfill
\hbox{\psfig{file=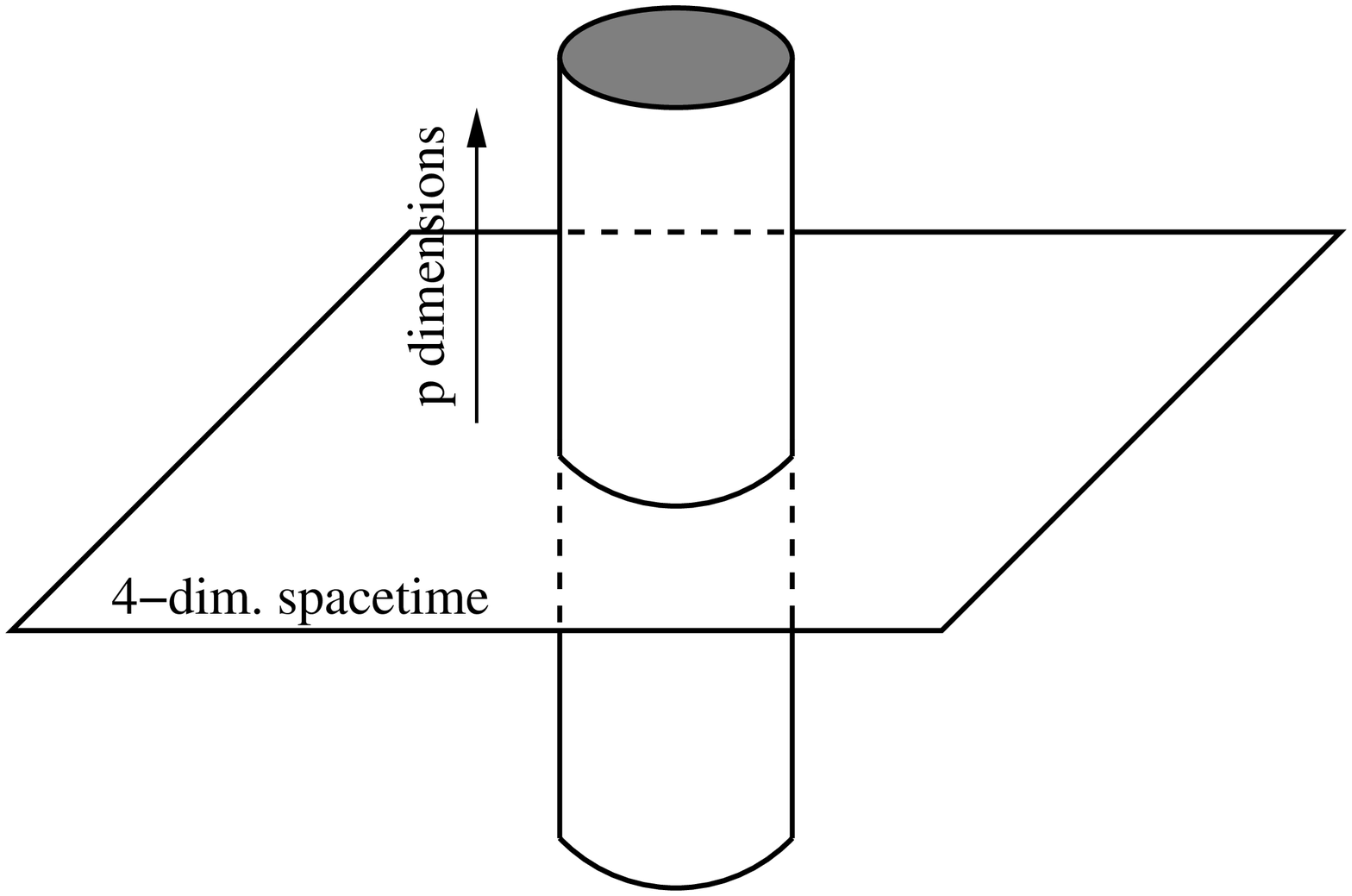,height=64mm}}\vfill}}
\vspace*{-20pt}
\caption{Schematic illustration of brane formation by super-Planckian
scattering of two incident particles with impact parameter $b$.}
\end{figure}

Let us focus attention on the Einsteinian brane \refb{brane.einstein3} and
compare the cross section for $p$-brane production to the BH cross section
\refb{cross-BH}. The ratio of $p$-brane and BH cross sections is
\be\label{ratio1}
\Sigma(s;n,p,v_p)\equiv{\sigma_{ij\to br}\over\sigma_{ij\to BH}} =
v_p^{-{2w\over n+1}}{G_{\rm Ein}(n,p)\over G_{\rm Ein}(n,0)}
\left({s\over s_{\star}}\right)^{{w-1\over n+1}}\,.
\ee
Since $w>1$ for any $n\,\ge p>0$, $\Sigma$ becomes larger for higher energy. At
fixed $n$ and $s$ the value of $\Sigma$ depends on the dimensionality of the
brane and on the size of the extra dimensions. Let us consider the scenario
with $m$ extra dimensions compactified on the $L$ scale and $n-m$ dimensions
compactified on the $L'$ scale (see Eq.~\refb{asymm-compact}). The $L$-size
dimensions can be identified with the small dimensions and are compactified at
about the Planck scale. If we assume that the $p$-brane wraps on $r$ $L$-size
dimensions ($r\le m)$ and on $p-r$ $L'$-size dimensions, the $p$-brane volume
$v_p$ is
\be
v_p=l^{r}
{l'}^{p-r}=l^{nr-mp\over
n-m}\left({M_{\rm Pl}\over M_{\star}}\right)^{2(p-r)\over n-m}\,.
\label{Vp}
\ee
Substituting Eq.\ \refb{Vp} in Eq.\ \refb{ratio1} we find
\be
\Sigma(s;n,m,p,r) =l^{-\alpha}
\left({M_{\rm Pl}\over M_{\star}}\right)^{-\beta}
{G_{\rm Ein}(n,p)\over G_{\rm Ein}(n,0)}
\left({s\over s_{\star}}\right)^{{w-1\over n+1}}\,,
\label{ratio2}
\ee
where
\be
\alpha={2(nr-mp)\over (n-m)(n-p+1)}\ge 0\,,\qquad
\beta={4(p-r)\over (n-m)(n-p+1)}\ge 0\,.
\label{alphabeta}
\ee
In TeV scale gravity, $M_{\rm Pl}/M_{\star}\approx 10^{14}$ ($10^{16}$) for
$M_{\star}\approx 100$ TeV ($1$ TeV). Since $0\le (w-1)/(n+1)\le 1$, the
$p$-brane cross section is suppressed w.r.t.\ spherically symmetric BH
cross section by a factor $\approx 10^{14\beta}$ ($10^{16\beta}$). The
largest cross section is obtained for $p=r$, i.e., when the $p$-brane is
completely wrapped on the small-size dimensions:
\be
\Sigma(s;n,m,p\le m) =
l^{-{2p\over
n-p+1}}{G_{\rm Ein}(n,p)\over G_{\rm Ein}(n,0)}
\left({s\over s_{\star}}\right)^{{w-1\over n+1}}\,.
\label{ratio3}
\ee
Assuming $L=L_\star$ the $p$-brane formation process dominates the BH formation
process. When the $p$-brane is wrapped on some of the large extra dimensions,
the $p$-brane cross section is suppressed w.r.t.~BH cross section. The ratio
$\Sigma$ slightly increases with the dimension of the brane. Therefore, in a
spacetime with $m$ fun\-da\-men\-tal-scale extra dimensions and $n-m$ large
extra dimensions a $m$-brane is the most likely object to be created. The cross
sections are dramatically enhanced if the $p$-brane wraps around dimensions
which are smaller than the fundamental scale. In ST $T$ dualities and mirror
symmetries usually set a lower bound on compactification dimensions in the weak
coupling regime of order $O(L_\star)$. However, it is not unreasonable to
assume, for instance, $L=L_{\star}/2$ or $L=L_{\star}/4$. \footnote{It should
be stressed that $L_\star$ denotes the fundamental scale rather than the
minimum distance. Therefore, compactifications with size (not too) smaller than
$L_\star$ are possible in principle. Analogy with elementary quantum mechanics
is illuminating. $\hbar$ denotes the scale of quantum processes in quantum
mechanics. However, processes with scale smaller than $\hbar$ are possible; a
harmonic oscillator satisfies $\Delta x\Delta p=\hbar/2$. In TeV-gravity
$L_\star$ denotes the scale of quantum gravity processes. By analogy,
compactifications with size $L=L_{\star}/2$ seem not to be unreasonable.}  In
this case the cross section may be enhanced by one or even two orders of
magnitude.

\begin{figure}
\centerline{\psfig{file=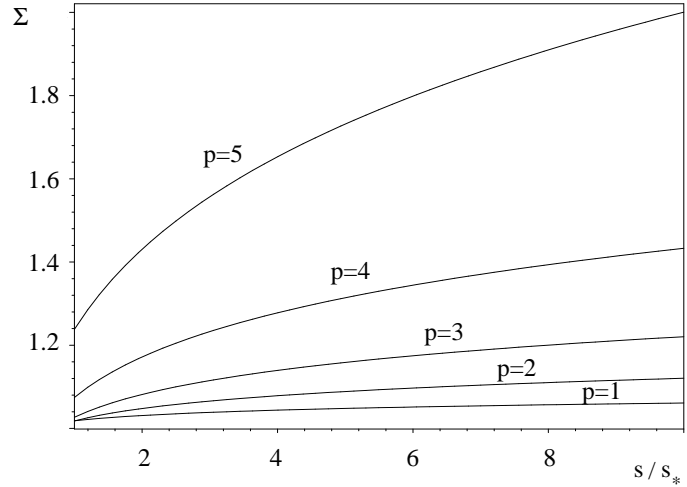,angle=270,width=10cm}}
\vspace*{8pt}
\caption{Ratio between the cross section for the creation of $p$-branes
($p\,\le\, m$) completely wrapped on fundamental-size dimensions and a
spherically symmetric BH in a 11-dimensional spacetime with $m=5$
fundamental-size extra dimensions $L=M_\star^{-1}=(100~{\rm TeV})^{-1}$ and
$n-m=2$ large extra dimensions of size $L'\approx 10^{12}~({\rm
TeV})^{-1}\gg L_{\star}$.}
\end{figure}

Let us consider the 11-dimensional spacetime as a concrete example. Assume
$m=5$ fundamental-scale extra dimensions $L=M_\star^{-1}=(100~{\rm TeV})^{-1}$
and two large extra dimensions $L\approx 10^{12}~({\rm TeV})^{-1}$. At
$s\approx 10 s_\star$ the cross sections for the formation of a 5-brane and a
4-brane completely wrapped on the fundamental-size dimensions are enhanced by a
factor $\approx 2$ and $\approx 3/2$ w.r.t.\ cross section for creation of a
spherically symmetric BH, respectively. (See Fig.~7). If the 5-brane wraps on
four extra dimensions with fundamental scale size and on one large extra
dimension, $\Sigma(s\approx 10 s_\star)$ is suppressed by a factor $\approx
10^{8}$. Assuming $L= L_{\star}/2$ ($L=L_{\star}/4$) the cross section for the
creation of $5$-branes in a $11$-dimensional spacetime is enhanced by a factor
$\approx 10$ (100). 

\begin{figure}
\centerline{\psfig{file=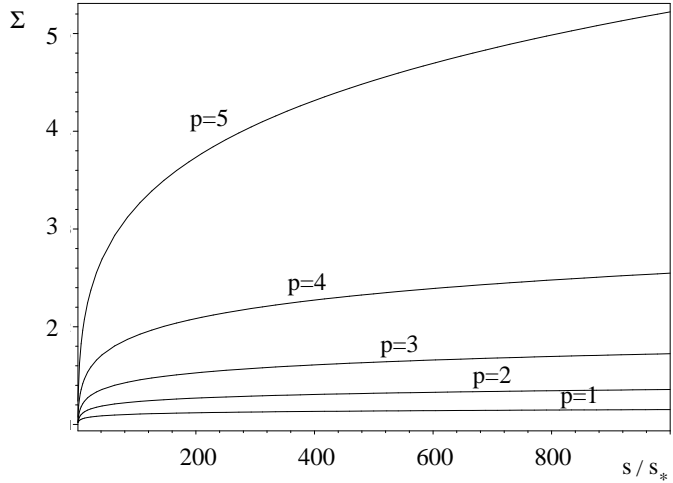,angle=270,width=10cm}}
\vspace*{8pt}
\caption{Ratio between the cross section for the creation of $p$-branes
($p\,\le\, m$) completely wrapped on fundamental-size dimensions and a
spherically symmetric black hole in a spacetime with $m=5$ fundamental-size
extra dimensions $L=M_\star^{-1}=(10~{\rm TeV})^{-1}$ and $n-m=2$ large extra
dimensions of size $L'\approx 10^{14}~({\rm TeV})^{-1}\approx 2\cdot
10^{-3}~{\rm cm}$. If the fundamental-size extra dimensions have size $L=
L_{\star}/4$ the cross sections are enhanced by a factor $\approx 100$, 16, 5,
2, 3/2 for $p=5$, 4, 3, 2 and 1, respectively.}
\end{figure}

Finally, let us consider symmetric compactifications. The ratio is
\be\label{ratio4}
\Sigma\approx
\left({M_{\rm Pl}\over M_{\star}}\right)^{-{4wp\over
n(n+1)}}{G_{\rm Ein}(n,p)\over G_{\rm Ein}(n,0)}
\left({s\over s_{\star}}\right)^{{w-1\over n+1}}\,.
\ee
In this case the cross section for $p$-brane formation is subdominant to the
cross section for BH formation. This result is understood qualitatively as
follows. If all the extra dimensions have (large) identical characteristic
size, the spacetime appears isotropic to the $p$-brane and a spherically
symmetric object is likely to form (see discussion in Sect.~3.1). Conversely,
when the compactification is asymmetric non-spherically symmetric objects are
more likely to be created. In an asymmetric model with $m$ small extra
dimensions the most likely $p$-brane to form is that with the highest symmetry
compatible with spacetime symmetries, i.e., a $m$-brane.

\begin{figure}
\centerline{\psfig{file=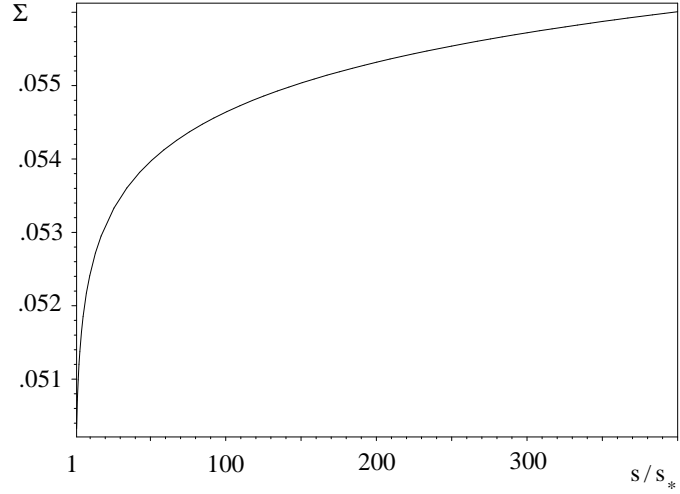,angle=270,width=10cm}}
\vspace*{8pt}
\caption{Ratio between the cross section for the creation of  a string ($p=1$)
and a spherically symmetric BH in a 11-dimensional spacetime with symmetric
compactification ($L_\star = 1$ TeV$^{-1}$). The cross section for string
formation is suppressed w.r.t.\ the cross section for BH formation by a factor
$\sim 1/20$.}
\end{figure}

\subsection{Brane Production on Fat Branes}
Recently, Cheung and Chou~\cite{Cheung:2002uq} have investigated Einsteinian
$p$-brane formation in FB and UED models (see Sect.~2.3.2). In these
models the production rate of $p$-branes relative to BHs can be enhanced by a
much larger factor than in the ADD scenario. If the SM particles propagate in some
of the extra dimensions, the cross section for producing a BH or a $p$-brane
must scale as $r_{p}^{k+2}$:
\be
\sigma_{ij\to br}^{(k)}=F_k(s)\pi r_p^{2+k}\,,
\ee
where $k\le n-2$ is the number of extra dimensions in which the SM particles
can propagate in addition to the four macroscopic ones.~\footnote{At least two
of the extra dimensions must be large to satisfy experimental constraints.
Therefore the SM particles can propagate in $n-2$ extra dimensions at the most.}
Setting $F_k(s)=1$ and assuming $L\sim L_\star$, the ratio of Einsteinian
$p$-brane and BH cross sections is given by
\be
\Sigma_k(s;n,m,p,r)\equiv{\sigma_{ij\to br}^{(k)}\over\sigma_{ij\to
BH}^{(k)}}\approx
\left({M_{\rm Pl}\over M_{\star}}\right)^{-\beta_k}
\left({G_{\rm Ein}(n,p)\over G_{\rm Ein}\gamma(n,0)}\right)^{k/2+1}
\left({s\over s_{\star}}\right)^{{w-1\over n+1}(k/2+1)}\,,
\label{ratio-fat}
\ee
where
\be\beta_k={2(2+k)(p-r)\over (n-m)(n-p+1)}\ge 0\,.
\label{alphak}
\ee
Equation \refb{ratio-fat} reduces to Eq.~\refb{ratio2} for $k=0$. For
$p$-branes completely wrapped on the small extra dimensions ($p=r$)
Eq.~\refb{ratio-fat} is
\be
\Sigma_k(s;n,p\le m)\approx
\left({G_{\rm Ein}(n,p)\over G_{\rm Ein}(n,0)}\right)^{k/2+1}
\left({s\over s_{\star}}\right)^{{w-1\over n+1}(k/2+1)}\,.
\label{ratio-fat2}
\ee
The largest ratio is obtained for $k=p=m$, {\it i.e.} when the SM particles
propagate in all small extra dimensions and the dimension of the brane is
equal to the number of small extra dimensions. Cheung and Chou have
computed the ratio \refb{ratio-fat2} for various values of $n$ and $p$ when
$k=p=m$. The highest ratio is obtained for a five-brane in a 11-dimensional
spacetime ($\Sigma\approx 105$). Therefore, in the FB or UED models the ratio of
$p$-brane production to BH production can be as large as $\sim 100$ without
requiring compactification radii smaller than the fundamental scale. The ADD
model with small extra dimensions $L\sim L_\star$ gives at most a ratio of
order $O(1)$.
\subsection{Brane Decay and Thermodynamics}
In contrast to BHs, the fate of $p$-branes depends strongly on the model.
Branes are extended objects endowed with tension. Therefore, they are unstable
unless some extra mechanism intervenes to stabilize them. A brane with horizon
and non-vanishing entropy evaporates by emitting thermal Hawking
radiation~\cite{Hawking:sw}. For instance, SUGRA branes of Sect.~4.2 evaporate
by Hawking radiation, provided that the Bogomol'ny bound is not
saturated~\cite{Stelle:1996tz,Stelle:nv}. The thermodynamics of generic branes
has been studied by Duff {\it et al.} in Ref.~\cite{Duff:1996hp}. However, a
thorough investigation of brane evaporation in the TeV scenario is still
missing.

The Hawking evaporation process does not occur for $p$-branes without horizon.
Although the decay process of singular branes is not understood, string field
theory
arguments~\cite{Sen:1999mh,Sen:1999mg,Sen:1999xm,Moriyama:2000dc,Rastelli:2000hv,Lee:2001cs,Lee:2001ey}
and analogy to cosmic
strings~\cite{Eardley:1995au,Hawking:1995zn,Gregory:1995hd} provide some clue
about brane decay. String field theory suggest that a higher-dimensional brane
can be seen as a lump of lower-dimensional branes. The tension of the brane
causes the latter to decay in lower dimensional branes, and eventually to
evaporate in gauge radiation. A bosonic non-supersymmetric brane can be
considered as an intermediate state in the scattering process. Pursuing further
the analogy with particle physics, BHs can be regarded as a metastable
particles and branes their resonances.~\footnote{The author is grateful to
Angela Olinto for this remark.}

The Einsteinian uncharged $p$-brane \refb{brane.einstein3} does not emit
Hawking radiation during the earlier stages of the decay because of the
singularity at $r=r_p$. However, the brane eventually decays in $0$-branes and
evaporate in a time of order $M_\star^{-1}$ by emitting ``visible'',
possibly thermal, brane and bulk quanta. Though the intermediate states of
the decay of a singular brane are highly dependent on the details of the theory
considered, we do not expect significant qualitative differences as far as the
final evaporation stage is concerned. It has also been conjectured that
singularities could explode in a sudden burst instead of evaporating via the
Hawking process~\cite{Iguchi:2001ya,Casadio:2001wh}. However, no information
about decay products nor estimate of the life-time of the singularity are
currently available. In either scenario (dimensional cascade followed by
Hawking evaporation {\it vs.}~naked singularity explosion) there is no reason
to believe that brane decay does not lead to production of visible particles. 
\subsection{Branes in the Randall-Sundrum Model}
Brane production can be easily accommodated in the RS scenario~\cite{rsbrane}.
The procedure follows closely that of Chamblin {\it et al.} for the black
string in AdS~\cite{Chamblin:1999by}. In conformal coordinates the
five-dimensional RS metric \refb{RS-metric} is
\be
ds^2={\lambda^2\over (|z|+z_0)^2}\left[\eta_{\mu\nu}\, dx^\mu dx^\nu+dz^2
\right]\,,
\label{warp-ads}
\ee
where $\lambda$ is the AdS radius and the (single) brane is located at $z=z_0$.
Motivated by ST/M-theory, in the following we consider a generalized
$D$-dimensional RS model with $D\ge 5$. One of the $n$ extra dimensions is
large and warped and provides the solution to the hierarchy problem. The
remaining $n-1$ extra dimensions are flat and are of order of the
fundamental Planck scale. The $D$-dimensional RS metric satisfies the Einstein
equations
\be
R_{ab}={D-1\over\lambda^2}g_{ab}=\Lambda g_{ab}\,.
\label{Einstein-ads}
\ee
If we replace the [$(D-1)$-dimensional] Minkowski metric $\eta_{\mu\nu}$ in
Eq.\ \refb{warp-ads} with any Ricci flat metric, the Einstein equations
\refb{Einstein-ads} are still satisfied. Therefore, we can substitute
Eq.~\refb{brane.einstein3} in Eq.~\refb{warp-ads} and obtain
\be
ds^2={\lambda^2\over (|z|+z_0)^2}\left[ds_{br}^2+dz^2\right]\,,
\label{pbrane-rs}
\ee
where $ds_{br}^2$ is the metric \refb{brane.einstein3}. Equation
\refb{pbrane-rs} describes a $p$-dimensional brane ($p\le D-5$) propagating on
the $(D-1)$-dimensional wall located at $z=z_0$. In the original
five-dimensional RS model the metric \refb{pbrane-rs} is
\be
ds^2={\lambda^2\over (|z|+z_0)^2}\left[-R(r)dt^2+R(r)^{-1}dr^2+r^2
d\Omega^2_{2}+dz^2\right]\,.
\label{bh-rs}
\ee
Equation \refb{bh-rs} describes a four-dimensional spherically symmetric
BH propagating on the wall at $z=z_0$.~\footnote{Alternatively, a black string
propagating in the full five-dimensional AdS spacetime.}. In eleven dimensions
the metric \refb{pbrane-rs} reads
\be
ds^2={\lambda^2\over (|z|+z_0)^2}\left[R^{{\Delta\over p+1}}(-dt^2+dz_i^2)+
R^{{1-\Delta\over
q-1}}\left(R^{-1}dr^2+r^2d\Omega^2_{8-p}\right)+dz^2\right]\,.
\label{11brane-rs}
\ee
where $\Delta=\left[(1-p/8)(p+1)\right]^{1/2}$ and $p\le 6$.

The cross section of brane production in the RS model depends on the AdS radius
and on the location of the background brane $z_0$. Since only the warped
dimension is large, the volume of a brane has Planck size. In the warped
model we expect larger cross sections compared to the ADD scenario: the maximal
dimension of the brane is $p=D-5$ in the former, while experimental constraints
limit $p$ to $D-6$ in the latter. A more complete investigation of brane
production in RS scenario is currently in progress~\cite{rsbrane}. The
phenomenology of RS BHs has been recently investigated in
Ref.~\cite{Anchordoqui:2002fc}
\section{Brane formation at Particle Colliders}
Alike BHs, $p$-brane formation could be observed at particle colliders. The
brane cross sections for proton-proton collisions and the brane production
rates at LHC have been computed by Ahn and the author in Ref.~\cite{Ahn:2002zn}
and independently by Cheung~\cite{Cheung:2002aq}. LHC with a proton-proton CM
energy of $14$ TeV will likely offer the first opportunity to observe brane
formation. The total cross-section for a proton-proton event is given by Eq.\
\refb{cross-pp} where $\sigma_{ij\to BH} \to \sigma_{ji\to br}$:
\be
\sigma_{pp\to br}(x_m;n,p,v_p)\sim\sum_{ij}\int_{x_m}^1dx\,\int_x^1 {dy\over
y}\,f_i(y,Q)f_j(x/y,Q)\,\sigma_{ij\to br}(xs;n,p,v_p)\,,
\label{cross-pp-br}
\ee
The cross sections for production of Einsteinian branes at LHC are plotted in
Fig 10. Assuming a fundamental Planck scale of $M_{\star}=2$ TeV and $D=10$
dimensions, the cross sections are in the range $10^{-4}-10^{3}$ pb and
increase for increasing brane dimension as expected. Therefore, the production
rate of higher-dimensional branes is higher than production rate of spherically
symmetric BHs. For a minimum brane mass of $M_{\rm min}=3$ TeV, the cross
section for a formation of a five- and a two-Einsteinian brane is
$\sigma_5\approx 250$ pb and $\sigma_2\approx 90$ pb, respectively. Therefore,
with a LHC luminosity ${\cal F}=3\cdot 10^{5}~{\rm pb}^{-1}~{\rm yr}^{-1}$
we expect a five-brane event and a two-brane event approximately every .5 and 1
second.  

\begin{figure}
\centerline{\psfig{file=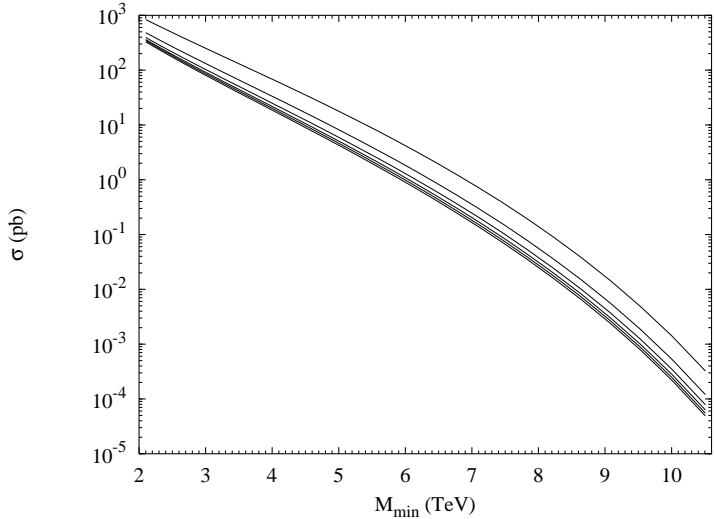,angle=270,width=10cm}}
\vspace*{8pt}
\caption{Cross sections (pb) for the formation of branes more massive than
M$_{\rm min}$ (TeV) at LHC ($D=10$, $p=0\dots 5$ from below). The volume of the
branes is assumed to be equal to one in fundamental Planck units ($M_\star=2$
TeV).}
\end{figure}

The production of branes at particle colliders - if observed - would allow the
investigation of the structure of the extra dimensions. Brane cross sections
are very sensitive to the size of the brane, which is related to the size of
the compactified extra dimensions around which the brane wraps. Since $v_p$ is
constant w.r.t.~the sum on partons, the total cross section is enhanced if the
length of the extra dimensions is sub-Planckian. For instance, the cross
section of a five-brane wrapped on extra dimensions with size 1/2 of the
fundamental scale is enhanced by a factor $\approx 10$. 

If supersymmetry is unbroken and SUGRA describes the physics at energies
above the TeV scale, high-energy particle scattering at particle collider could
produce charged SUGRA branes (see Sect.~4.2). In order to produce SUGRA branes
the collision must be either of the two processes
\begin{eqnarray}
&&p^+~+~p^+~\to~{\rm brane}~+~X^q\,,\nonumber\\
&&p^+~+~p^+~\to~{\rm brane}~+~{\rm antibrane}\,,\nonumber
\end{eqnarray}
where $X^q$ denotes a set of particles with total charge equal to $2e^{+}$
minus the brane charge. The cross sections for production of SUGRA branes are
comparable to those of Einsteinian branes. (See, e.g., Fig.~11). 

The experimental signatures depend on the decay process. If the brane possesses
a horizon or decays into lower-dimensional branes, we expect an observed
hadronic to leptonic activity of roughly 5:1, high multiplicity and emission of
a few hard visible quanta at the end of the process. However, if the brane 
explodes in a sudden burst, or a stable SUGRA brane forms, the experimental
signatures could be drastically different. In particular, an extremal brane
would be detected either as missing energy or as a stable charged heavy
particle. A comprehensive study of the experimental signatures of brane
production in particle colliders is currently missing.

\begin{figure}
\centerline{\psfig{file=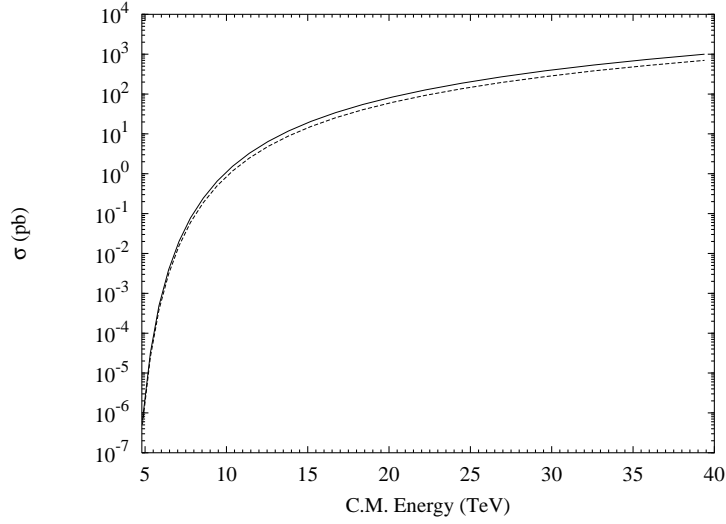,angle=270,width=10cm}}
\vspace*{8pt}
\caption{Cross sections for creation of magnetic (solid curve) and electric
(dashed curve) 11-dimensional SUGRA branes (Eqs.~\refb{brane-sugra-magn} and
\refb{brane-sugra-el}) by proton-proton scattering with $M_\star=2$ TeV,
minimum mass $M_{\rm min}=2M_\star$, and unit volume. The cross sections for
the corresponding Einsteinian two- and five-brane are enhanced by a factor of
$\sim 2.2$ and $\sim 2.7$, respectively.}
\end{figure}

\section{Brane Production and High Energy Cosmic Rays}
In the TeV scenario UHECR can create $p$-branes. Alike BHs (see Sect.3.3) the
production and the subsequent decay of $p$-branes in the terrestrial atmosphere
could lead to formation of showers that might be detected by cosmic ray
experiments~\cite{Nagano:ve}. The best candidate for brane formation are
cosmogenic neutrinos~\cite{Engel:2001hd}. The total cross section for neutral
Einsteinian brane production was first computed in Ref.~\cite{Jain:2002kf} and
is obtained from Eq.~\refb{cross-nup} by substituting $\sigma_{ij\to BH}$ with
$\sigma_{ji\to br}$:
\be
\sigma_{\nu p\to br}(x_m;n,p,v_p)\sim\sum_{i}\int_{x_m}^1dx\,
f_i(x,Q)\,\sigma_{ij\to br}(xs;n,p,v_p)\,,
\label{cross-nup-br}
\ee

\begin{figure}
\centerline{\psfig{file=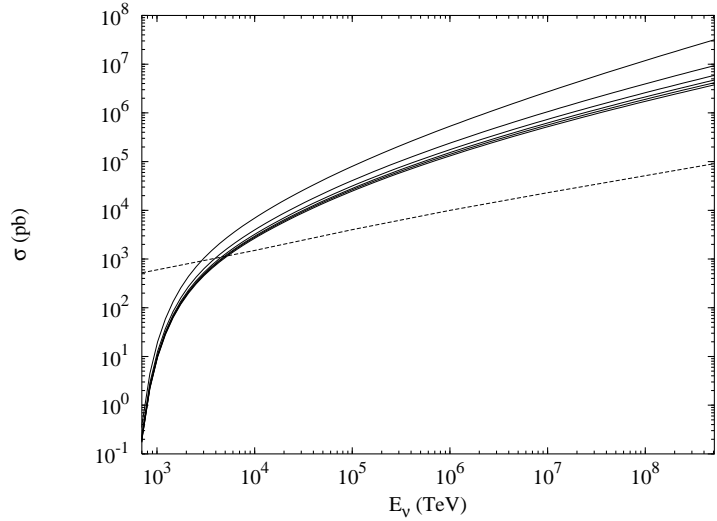,angle=270,width=10cm}}
\vspace*{8pt}
\caption{Cross sections for creation of neutral Einsteinian branes by UHECR
scattering with $M_\star=M_{\rm min}=1$ TeV, unit volume, and $n=6$ ($p=0\dots
5$ from below). The dashed curve is for the SM process.}
\end{figure}

As usual, the cross section is enhanced if the small extra dimensions are
compactified below the fundamental scale. The total cross section for a
neutrino with energy $\sim 10^{9}$ TeV might reach the 100 mb range, a value
which is comparable to SM nucleon-nucleon cross section at these energies. This
fact led Jain {\it et al.}~\cite{Jain:2002kf} to conjecture that brane
formation could account for the observed air showers above the GZK
cutoff~\cite{Greisen:1966jv,Zatsepin:1966jv}. Their analysis was refined by
Anchordoqui {\it et al.}~in Ref.~\cite{Anchordoqui:2002it} who found that cross
sections of order $\sim$ 100 mb are obtained only for $n=1$ or $2$, small
compactification radii, and small $M_\star$, a region of the moduli space which
is excluded by experimental constraints (see Sect.~2.4). Therefore, it seems
unlikely that brane formation by cosmogenic neutrinos in flat compactification
models may resolve the GZK paradox. On the contrary, scenarios with warped
dimensions still allow for an explanation of super-GZK events. 

The current non-observation of $p$-brane events puts stringent limits on the
fundamental scale $M_\star$. Neutrino cross sections \refb{cross-nup} of order
of a mb may enhance deep quasi-horizontal shower rates. The number of deep
showers per unit time is~\cite{Anchordoqui:2002it}
\be
N=\int dE_\nu N_A {d\Phi\over dE_\nu}\sigma_{ij\to br}(E_\nu)A(E_\nu)\,,
\label{horiz-rate}
\ee
where $N_A$ is Avogadro's number, $d\Phi/dE_\nu$ is the neutrino flux,
$A(E_\nu)$ is the acceptance for quasi-horizontal showers in cm$^3$ water
equivalent steradians. Anchordoqui {\it et al.} have computed the event rate  
\refb{horiz-rate} for the AGASA experiment \cite{AGASA} and compared the result
to actual data. In $\sim 1710$ days of data taking, AGASA found a single deep
horizontal event with an expected background of 1.72 \cite{Yoshida:2001},
leading to a 95\% c.l.~limit of 3.5 events for brane creation. This result sets
bounds both on the fundamental scale and on the size of small extra dimensions.
For a number of large extra dimensions $n-m\ge 4$ absence of deep
quasi-horizontal showers sets the most stringent bounds on the fundamental
scale $M_\star$. For $(n,m)=(6,5)$ and  $(n,m)=(7,5)$ the lower bound on
$M_\star$ ranges from $\sim 2-3$ TeV ($L\sim L_\star/2$) to $\sim 0.7$ TeV
($L\sim 10 L_\star$). Non-observation of brane cascades at the Auger
observatory \cite{Auger} will set even more stringent bounds on $M_\star$.

Finally, Sigl~\cite{Sigl:2002bb} has recently given constraints on the moduli
space of BH and $p$-brane cross sections using UHECR and neutrino data. Since
brane cross sections grow slower than $s$ (see Eq.~\refb{cross-brane}) the
moduli space is not strongly constrained at energies above the TeV scale. 
\section{TeV Branes and Cosmology}
In the standard cosmological scenario~\cite{Kolb:Book} and in the new
brane-world cosmological models~\cite{Khoury:2001wf}, the temperature of the
early universe is expected to have exceeded TeV values. Therefore, creation of
$p$-branes could have been a common event in the early universe. At
temperatures $\tau$ above the fundamental scale we expect a plasma of branes 
with mass $M_p\sim\tau$ in thermal equilibrium with the primordial bath. At
temperatures of order of TeV, branes decouple from the thermal plasma. If the
branes are long-lived or stable, $p$-brane relics would appear to an observer
today like heavy (supersymmetric?) particles with mass $M_p\sim$ TeV and cross
sections $\sigma_{br}\laq$ pb, thus providing a candidate for dark matter. The
presence of a gas of branes could also solve the initial singularity and
horizon problems of the standard cosmological model without relying on an
inflationary phase~\cite{Alexander:2000xv,Brandenberger:2002kj}.  
\section{Conclusion and Outlook}
In TeV gravity theories, processes at energies $\gaq$ TeV may experimentally
test quantum gravitational effects. Particle collisions with CM energy larger
than $M_\star$ and sufficiently small impact parameter generate NPGOs: BHs,
string balls and branes. Formation and subsequent decay of super-Planckian NPGO
should be detectable in future particle and UHECR experiments. In cosmology,
primordial creation of BHs and branes could have played an important role in
the dynamics of the very early universe. Brane relics could be the dark matter
which is observed today. 

BH formation in LED models has attracted a lot of attention in the scientific
community. The investigation of brane production is just at the beginning. Up
to now, studies on brane production have focused essentially on the computation
of cross sections. Brane cross sections are comparable or dominant to BH cross
sections in spacetimes with flat asymmetric compactifications, fat branes, and
ST models. BH factories are also brane factories. On the other hand, the
non-observation of BH and brane events with current UHECR detectors sets lower
bounds on the fundamental scale and on the structure and size of compactified
dimensions.~\footnote{Unlike BHs, $p$-branes which do not possess an event
horizon, such as \refb{brane.einstein3}, do not cloak hard processes. Different
hard super-Planckian processes lead to different experimental signatures which
depend on the physics of the collision and on the structure of the extra
dimensions.} Eventually, either BH and brane creation will be discovered in
particle collider and UHECR experiments or the fundamental scale $M_\star$ will
be pushed so high to make low-scale gravity models worthless. 

The investigation of the theoretical and phenomenological properties of brane
formation is presently limited by the ignorance of QG and the physics of
gravitational collapse. Poor understanding of brane formation and decay and
lack of an accurate computation of brane cross sections (both from theoretical
and numerical points of view) are amongst the unsolved theoretical issues. The
future studies on phenomenological implications of brane production should
focus on experimental signatures in particle collider and UHECR experiments. In
particular, a thorough investigation of the experimental differences between BH
and brane production at particle colliders (see Sect.~3.2) is still missing.
The study of cosmological and astrophysical implications of $p$-brane
production (early universe physics, high-energy cosmic rays) is also of primary
importance. BH and brane formation could play a fundamental role in all
physical processes with energy above the TeV scale.
\section*{Acknowledgements}
I am very grateful to E.-J.~Ahn, B.~Harms and A.~Olinto and for interesting
discussions and fruitful comments, and to I.~Antoniadis, U.~d'Alesio, G.~Dvali,
J.~Feng, S.~Giddings, C.~Giunti, H.~Goldberg, R.~Gregory, A.~Hanany,
G.~Karatheodoris, S.~Pinzul, J.~Polchinski, Y.~Uehara, A.~Vilenkin and
B.~Zwiebach for useful comments and correspondence. This work is supported in
part by funds provided by the U.S.\ Department of Energy under cooperative
research agreement DE-FC02-94ER40818. I thank the Department of Astronomy and
Astrophysics of the University of Chicago for kind hospitality.
\appendix
\section*{Appendix}
We summarize other definitions of the fundamental scale that have been used in
the literature.
\subsection*{Giudice-Rattazzi-Wells (GRW) Notations}
The observed Planck mass $M_{\rm Pl}$ is defined as in Eq.~\refb{MPl}. The
fundamental Planck mass is $M_D=[8\pi/(2\pi)^n]^{-{1\over n+2}}M_\star$. For a
$n$-dimensional symmetric toroidal compactification with radii $R$ the
fundamental Planck scale is $M_D=(8\pi G_4 R^n)^{-{1\over n+2}}$, where $L=2\pi
R$ is the length of the extra dimensions. The $D$-dimensional Newton constant
$G_D$ is defined as $G_D=(2\pi)^n/(8\pi M_D^{2+n})$. GRW notations are used in
Refs.~\cite{Giudice:1998ck,Anchordoqui:2001cg,Cheung:2002aq,Acciarri:1999kp,Acciarri:1999jy}.
\subsection*{Cullen-Perelstein (CP) Notations}
The $D$-dimensional Newton's constant $G_D$ is defined as in Eq.~\refb{G4}. The
fundamental Planck mass $M_D$ is defined as $M_D=[(2\pi)^{D-4}/(4\pi
G_D)]^{1\over D-2}$. For symmetric toroidal compactifications with radii $R$
$M_D=(4\pi G_4 R^n)^{-{1\over n+2}} = [4\pi/(2\pi)^n]^{-{1\over n+2}}M_\star$.
CP notations are used in
Ref.~\cite{Giddings:2001bu,Peskin:2000ti,Cullen:2000ef,Hannestad:2001jv,Hannestad:2001xi,Fairbairn:2001ct,Hall:1999mk}.
\subsection*{Han-Lykken-Zhang (HLZ) Notations}
The relation between the observed Planck mass and the fundamental scale $M_s$
is $M_{\rm Pl}^2=\Omega_{n-1}(2\pi)^{-n}V_nM_s^{n+2}$, where
$\Omega_{n-1}=2\pi^{n/2}/\Gamma(n/2)$ is the volume of the unit sphere in $n-1$
dimensions. For a symmetric toroidal compactification the previous relation
simplifies to $M_{\rm Pl}^2=\Omega_{n-1}R^nM_s^{n+2}$. The relation between $M_s$
and $M_\star$ is $M_s=[\Omega_{n-1}/(2\pi)^n]^{-{1\over n+2}}M_\star$. HLZ
notations are used in Refs.~\cite{Han:1998sg,Abbott:2000zb,Barger:1999jf}.
\subsection*{EOT-WASH Group Collaboration Notations}
In Refs.~\cite{Hoyle:2000cv,Adelberger:2002ic} the fundamental
Planck mass $M^\star$ is defined by the relation
\be
R^\star={1\over M^\star}\left({M_{\rm Pl}\over M^\star}\right)^{2/n}\,,
\ee
where $R^\star$ is the radius of the symmetric compactification. For a toroidal
compactification, setting $R^\star=L/(2\pi)$, we find the following relations:
\be
M^\star=(2\pi)^{n\over n+2}M_\star=(8\pi)^{1\over n+2}M_D\,.
\ee

\end{document}